\documentclass[aps,pra,twocolumn,groupedaddress]{revtex4-1}
\usepackage{amsmath,amssymb,bm,graphicx}
\newcommand{\ii}{{\mathrm{i}}}
\newcommand{\ee}{{\mathrm{e}}}
\newcommand{\dd}{{\mathrm{d}}}

\begin{document}
\title{Electrostatics of optical rectification in metallic particles}
\author{Tetsuyuki OCHIAI}
\affiliation{Research Center for Functional Materials, National Institute for Materials Science}
\date{\today}

\begin{abstract}
We present an electrostatic theory of the optical rectification, namely, the static photovoltage or photocurrent generation under a light illumination, in metallic particles.  
The hydrodynamical model for the charge carriers in the metals is employed. By solving the hydrodynamic equation and Maxwell equation perturbatively, we obtain analytically the second-order susceptibility, from which the optical rectification is explained.  
Electrostatic potential problems involved  in the optical rectification under the local response approximation are formulated in arbitrary geometries, and then are solved for simple geometries of metallic planar interfaces, slabs, cylinders, and spheres. The photovoltage and photocurrent spectra, their incident-angle  dependence, and the electrostatic potential distribution for a  incident plane-wave light are demonstrated and discussed in a context of  plasmonic resonances.  
\end{abstract}

\pacs{}
\maketitle

\section{Introduction}

Plasmon resonances are sources of many interesting phenomena in physics and engineering \cite{Raether-plasmon-book}. 
Highly concentrated radiation fields of the plasmon resonances cause strong light-matter interactions, yielding enhancements of the Raman scattering, florescence, nonlinear optics, and so on. 
So far, various types of  enhanced nonlinear optical processes due to plasmons have been investigated \cite{kauranen2012nonlinear}.  
 
Here, we focus on the optical rectification (OR) \cite{bass1962optical},  which is a static voltage and/or current generation under a light illumination  as a second-order nonlinear-optical process.  
The OR is less common in nonlinear optics compared to, for instance, the second-harmonic generation (SHG) and Kerr effect. However, its importance is obvious in ultrafast photodetection, terahertz light source, solar cell, and wireless communication.  
Moreover, there have been renewed interests in anomalous photovoltage, or in other words, the shift current, in semiconductors and insulators \cite{nakamura2017shift,cook2017design}. There, topological effects play crucial roles. To differentiate among various types of photovoltage generation mechanisms, a clear understanding of the OR is in order.

Metal is a remaining and intriguing platform of the OR, because the plasmon resonances can affect strongly the OR.  In addition, metal is theoretically rather simple. It can be described semi-analytically by the Drude response. 
So far, detailed investigations of plasmon-enhanced OR have been made in homogeneous metallic films \cite{vengurlekar2005surface,PhysRevB.84.035447,PhysRevLett.123.053903, khichar2021new}, nanowire \cite{PhysRevLett.103.186801}, one- and two-dimensional diffraction gratings  \cite{PhysRevLett.103.103906,english2011hydrodynamic,kurosawa2012surface,noginova2013plasmon,PhysRevA.95.033844,piltan2017optical,Moroshkin-paper}, and porous metals \cite{akbari2015photo}.  
In addition, plasmon-enhanced ORs in gated graphene by grating couplings have been investigated \cite{PhysRevB.93.075422,ochiai2017enhanced}.

In most investigations, theoretical understanding of the OR is based on an intuitive picture of the radiation force acting on the metal carriers.  
In contrast, Goff and Schaich proposed a hydrodynamic approach to metal surfaces \cite{PhysRevB.56.15421}.  
Kurosawa et al  developed the hydrodynamical approach combined with  first-principles calculations of the Maxwell equation for photonic nanostruructures \cite{PhysRevA.95.033844}. Their study goes beyond the naive picture and clarifies the roles of the nonlocality in optical responses that are neglected in the naive picture.

However, still many important questions on the OR remain unsolved. 
How does the boundary affect the OR?  How do the voltage, electric field, internal current, and static polarization behave inside the specimen? 
Actual measurements of photovoltage are made by probing at boundaries of a finite-size specimen, so that sample edges may affect strongly the result.  
Also, there is a fundamental issue in polarity of the photocurrent, as raised recently by Strait et al \cite{PhysRevLett.123.053903}.

To answer these questions, 
We here formulate a rigorous theory of the OR within the local-response approximation (LRA) but dealing with metal boundaries analytically.  We first present an electrostatic theory of the OR in arbitrary geometry and apply it to simple geometries of planar interfaces, slabs, cylinders, and spheres. These structures have well-defined boundaries and semi-analytic treatment of optical responses are available. 
By solving the electrostatic potential problem explicitly, we can know how the current and voltage behave inside the finite-size specimens, and how the boundaries affect the results.   
Although nonlocal responses are often critical, we here restrict ourselves to the LRA and focus on possible geometrical effects taking account of the boundaries, and on the  field distributions. Detailed investigation of the interplay between geometric and nonlocal effects is beyond the scope of the present paper.

This paper is organized as follows. In Sec. II, we present an electrostatic theory of the OR in arbitrary geometries of metals using the hydrodynamical model of the metal carriers. Sections III, IV, V, VI  are devoted to study  planar interface, slab, cylinder, and sphere  geometries, respectively. We present the results of OR spectra and potential distributions in each systems.  
Finally, in Sec. VII, summary and discussion are given.

\section{Hydrodynamic theory}

Suppose we have a metallic system illuminated by a continuous-wave light.  
We employ the hydrodynamical approach to the charge carriers in the metallic system \cite{PhysRev.174.813,PhysRevB.56.15421}.  
The hydrodynamical equation for the carriers with mass $m$ and charge $e$ is given by 
\begin{align}
m\left(\frac{\partial{\bm v}}{\partial t}+({\bm v}\cdot{\bm \nabla}){\bm  v}+\gamma{\bm v}\right)=e({\bm E}+{\bm v}\times{\bm B}),  
\end{align}
where ${\bm v}$ is the velocity field, $\gamma$ is a phenomenological damping factor,  ${\bm E}$ and ${\bm B}$ are the electric and magnetic fields, respectively. 
Here, we neglect the pressure term as the LRA \footnote{In the hydrodynamic approach,  the Fermi degeneracy pressure term is usually included, giving rise to nonlocal optical responses. }.     
The charge density $\rho$ and current density  ${\bm j}=\rho {\bm v}$ satisfy the charge conservation law as  
\begin{align}
\frac{\partial \rho}{\partial t}+{\bm \nabla}\cdot{\bm j}=0.  
\end{align}

Starting from the equilibrium charge density $\rho^{(0)}({\bm x},t)=\rho_0(=en_0)$ and current density ${\bm j}^{(0)}({\bm x},t)=0$, together with the incident radiation fields of ${\bm F}^{(1);\mathrm{inc}} \; ({\bm F}={\bm E},{\bm B})$ with angular frequency $\omega$, we solve the hydrodynamic equation and Maxwell equation perturbatively. 
The Maxwell equation to be solved is given by 
\begin{align}
&{\bm \nabla}\cdot(\epsilon_0\epsilon_{\infty}{\bm E})=\rho,\\
&{\bm \nabla}\cdot{\bm B}=0,\\
&{\bm \nabla}\times{\bm E}=-\frac{\partial {\bm B}}{\partial t},\\
&{\bm \nabla}\times{\bm B}=\mu_0{\bm j}+\frac{\partial (\epsilon_\infty{\bm E})}{c^2\partial t}, 
\end{align}
inside the metal, where $\epsilon_\infty$ is the permittivity due to ions and bounded electrons, which is assumed to be frequency independent.  Outside the metal, the Maxwell equation is obtained by putting $\epsilon_{\infty}\to \epsilon_\mathrm{out}$ being $\epsilon_\mathrm{out}$ the permittivity of the outer medium, and $\rho,{\bm j}\to 0$.

We expand various fields in a power series of the amplitude of the incident light:  
\begin{align}
&\rho=\rho^{(0)}+\rho^{(1)}+\rho^{(2)}+\cdots,\\
&{\bm j}={\bm j}^{(0)}+{\bm j}^{(1)}+{\bm j}^{(2)}+\cdots,\\
&{\bm F}={\bm F}^{(0)}+{\bm F}^{(1)}+{\bm F}^{(2)}+\cdots
\end{align}
Here, we assume that $\rho^{(0)}$ of the carriers is canceled with the background charge density of the ions and bounded electrons, so that ${\bm F}^{(0)}=0$.

In the first order, we have the following  linear response: 
\begin{align}
&\ddot{\bm P}^{(1)}+\gamma \dot{\bm P}^{(1)}=\frac{e\rho_0}{m}{\bm E}^{(1)},\\
&\rho^{(1)}=-{\bm \nabla}\cdot{\bm P}^{(1)}, \quad {\bm j}^{(1)}=\frac{\partial {\bm P}^{(1)}}{\partial t}. \label{Eq_linear}
\end{align}
The first-order radiation field is time-harmonic as  
\begin{align}
&{\bm F}^{(1)}({\bm x},t)=\Re[\tilde{\bm F}_\omega^{(1)}({\bm x})\ee^{-\ii\omega t}], 
\end{align}
where $\Re$ stands for the real part. 
In terms of the complex field, we obtain 
\begin{align}
-\omega(\omega +\ii \gamma)\tilde{\bm P}_\omega^{(1)}=\frac{e\rho_0}{m}\tilde{\bm E}_\omega^{(1)}. 
\end{align}
The linear response results in the Drude permittivity as 
\begin{align}
&{\bm D}=\epsilon_0\epsilon_\infty{\bm E}+{\bm P},\\
&\tilde{\bm D}_{\omega}^{(1)}=\epsilon_0\epsilon(\omega)
\tilde{\bm E}_{\omega}^{(1)},\\
&\epsilon(\omega)=\epsilon_\infty -\frac{\omega_\mathrm{p}^2}{\omega(\omega +\ii \gamma)}, \label{Eq_Drude}\\
&\omega_\mathrm{p}^2=\frac{e\rho_0}{\epsilon_0 m}. 
\end{align}
We can, in principle, solve the Maxwell equation with the Drude  permittivity,  by employing appropriate methods.

In the second order, the current density is composed of the direct current (DC) and alternating current of angular frequency $2\omega$: 
\begin{align}
{\bm j}^{(2)}({\bm x},t)={\bm j}_\mathrm{DC}^{(2)}({\bm x})+\Re[\tilde{\bm j}_{2\omega}^{(2)}({\bm x})\ee^{-2\ii\omega t}].
\end{align} 
The former current is responsible to the OR and the latter is to the SHG.   
Here, we focus on the direct current.

The direct current  is explicitly given by 
\begin{align}
&{\bm j}_\mathrm{DC}^{(2)}=\frac{e\rho_0}{m\gamma}{\bm E}_\mathrm{DC}^{(2)}+{\bm i}_\mathrm{DC}^{(2)}, \label{Eq_DCcurrent}\\
&[{\bm i}_\mathrm{DC}^{(2)}]_i=\frac{e\epsilon_0\omega_\mathrm{p}^2}{2m\gamma\omega}\left(
-\Re\left[ \frac{1}{\omega -\ii \gamma}(\tilde{\bm E}_\omega^{(1)})_j^*\partial_i(\tilde{\bm E}_\omega^{(1)})_j\right]\right. \nonumber\\ 
&\hskip60pt \left. +\frac{\gamma}{\omega^2+\gamma^2} \partial_j\Im[(\tilde{\bm E}_\omega^{(1)})_i^*(\tilde{\bm E}_\omega^{(1)})_j].
\right)
\end{align}
In the first term of Eq. (\ref{Eq_DCcurrent}), $e\rho_0/({m\gamma})$ is the DC conductance and ${\bm E}_\mathrm{DC}^{(2)}$ is the DC electric field to be determined self-consistently. This term is an Ohmic current.  
The term ${\bm i}_\mathrm{DC}^{(2)}$ corresponds to the electromotive force (EMF) due to the second-order optical nonlinearity. 
Here, symbol $\Im$ stands for the imaginary part and the Einstein summation  convention is assumed for repeated indices.

Since the direct current is divergence free and the DC electric field is longitudinal, namely, 
\begin{align}
&{\bm \nabla}\cdot{\bm j}_\mathrm{DC}^{(2)}=0,\\
&{\bm E}_\mathrm{DC}^{(2)}=-{\bm \nabla}\phi_\mathrm{DC}^{(2)}, 
\end{align}
we obtain  the Poisson equation for $\phi_\mathrm{DC}^{(2)}$ as 
\begin{align}
&\Delta \phi_\mathrm{DC}^{(2)} = -\frac{1}{\epsilon_0\epsilon_ \infty}\rho_\mathrm{DC}^{(2)}, \label{Eq_poisson}\\
&\rho_\mathrm{DC}^{(2)}=-\frac{\epsilon_\infty\gamma}{\omega_\mathrm{p}^2}{\bm \nabla}\cdot{\bm i}_\mathrm{DC}^{(2)}. \label{Eq_chargedensity}
\end{align}
The problem thus reduces to the electrostatic potential one under a given charge density 
$\rho_\mathrm{DC}^{(2)}$. 
Equation (\ref{Eq_chargedensity}) implies a nontrivial relation between the 
static polarization and EMF current as 
\begin{align}
{\bm P}_\mathrm{DC}^{(2)}=\frac{\epsilon_\infty\gamma}{\omega_\mathrm{p}^2}{\bm i}_\mathrm{DC}^{(2)}.
\end{align}

To solve the potential problem, the boundary condition plays a crucial role. 
The boundary condition for the photovoltage calculation is the open-circuit condition, namely, 
\begin{align}
({\bm j}_\mathrm{DC}^{(2)})_\perp=0, \label{Eq_normalcurrent} 
\end{align}
at the metal surface. Here, subscript $\perp$ represents the surface normal component. 
This yields a Neumann-type boundary condition as 
\begin{align}
&\partial_\perp\phi_\mathrm{DC}^{(2)}=
\frac{\gamma}{\epsilon_0\omega_\mathrm{p}^2}({\bm i}_\mathrm{DC}^{(2)})_\perp.
\end{align}
Then, the photovoltage $V$ measured at two points on the surface is given by 
\begin{align}
V=\phi_\mathrm{DC}^{(2)}({\bm x}_1)-\phi_\mathrm{DC}^{(2)}({\bm x}_2).
\end{align}
On the other hand, the short-circuit current is evaluated as Eqs. (\ref{Eq_DCcurrent}) and (\ref{Eq_poisson}) by imposing 
\begin{align}
\phi_\mathrm{DC}^{(2)}({\bm x}_1)=\phi_\mathrm{DC}^{(2)}({\bm x}_2)
\end{align}
where ${\bm x}_1$ and ${\bm x}_2$ are connected by a zero-resistance wire.  
This is a Dirichlet-type boundary condition. 
By solving the Poisson equation Eq. (\ref{Eq_poisson}) with appropriate boundary condition, we can obtain the potential and photocurrent distributions in the specimen.

The OR is often argued intuitively by a momentum transfer or the radiation force. 
If we send light to a specimen,  the radiation force acts on the charge carriers in the specimen. The force induces a certain spatial asymmetry in the charge distribution, so that a nonzero static electric field is induced. 
In the open-circuit condition, the balance between the radiation force and Coulomb force results in a nonzero photovoltage. 

This intuitive argument is formulated by the Maxwell's stress tensor $T_{ij}$ as 
\begin{align}
&[{\bm j}_\mathrm{DC}^{(2)}]_i=\frac{e}{m\gamma}\left[\rho_0 ({\bm E}_\mathrm{DC}^{(2)})_i +\partial_j T_{ij} -2\partial_j K_{ij}\right], \label{Eq_DCforce}\\
&T_{ij}=\frac{1}{2}\Re\left[\epsilon_0\epsilon_{\infty}(\tilde{\bm E}_\omega^{(1)})_i^*(\tilde{\bm E}_\omega^{(1)})_j+
\frac{1}{\mu_0}(\tilde{\bm B}_\omega^{(1)})_i^*(\tilde{\bm B}_\omega^{(1)})_j \right. \nonumber \\ 
&\hskip50pt \left. -\frac{1}{2}\delta_{ij}\left(\epsilon_0\epsilon_{\infty}|\tilde{\bm E}_\omega^{(1)}|^2+\frac{1}{\mu_0}|\tilde{\bm B}_\omega^{(1)}|^2\right) \right],\\
&K_{ij}=\frac{1}{4}mn_0\Re\left[(\tilde{\bm v}_\omega^{(1)})_i^*(\tilde{\bm v}_\omega^{(1)})_j \right], \quad \tilde{\bm v}_\omega^{(1)}=-\ii \frac{\omega}{\rho_0}\tilde{\bm P}_\omega^{(1)}.
\end{align} 
Here, Eq. (\ref{Eq_DCforce}) is an alternative expression of Eq. (\ref{Eq_DCcurrent}). 
In a simple geometry such as planar surfaces, the current conservation ${\bm \nabla}\cdot{\bm j}_\mathrm{DC}^{(2)}=0$ gives Eq. (\ref{Eq_normalcurrent}) not only at the boundary but also in the bulk \cite{PhysRevB.56.15421}. 
In this case, we have  
\begin{align}
-\rho_0\partial_\perp\phi_\mathrm{DC}^{(2)}+\partial_jT_{\perp j}-2\partial_jK_{\perp j}=0.  
\end{align}
 This equation determines the potential distribution through the Maxwell stress tensor, which represent the radiation force acting on the charge carriers, and kinetic tensor $K_{ij}$.   By integrating the above equation in the metal region $\mathcal{V}$, we have 
 \begin{align}
 \rho_0\int_\mathcal{V} \dd^3x \partial_\perp\phi_\mathrm{DC}^{(2)}=
 \int_{\partial\mathcal{V}}\dd S_j(T_{\perp j}-2K_{\perp j}),  \label{Eq_forceint}
 \end{align}
 where $\partial\mathcal{V}$ represents the boundary surface of $\mathcal{V}$. 
 Since the carriers are bounded in the metal region, $(\tilde{\bm v}_\omega^{(1)})_\perp$ almost vanishes at the metal surface. 
 Therefore, the kinetic tensor part is negligible in Eq. (\ref{Eq_forceint})
 and the photovoltage is given by the stress tensor. 
 However, such a simplification is available only in limited cases. Therefore, in general, the radiation force does not fully explain the OR.

\section{Planar interface} 
Let us first consider a planar interface between a semi-infinite metal and air. The interface is normal to the $z$ direction and is taken to be at $z=0$ ($z<0$ corresponds to the metal).    
The incident plane wave  ${\bm E}^{(1);\mathrm{inc}}={\bm e}_0\exp(\ii{\bm k}\cdot{\bm x})$ 
is coming from the air side ($z>0$) and the photovoltage is induced in the $z$ direction. The photocurrent is also induced in the horizontal direction that is assumed to be infinitely extended.

In the first order, the radiation field in the metal is written as 
\begin{align}
&\tilde{\bm E}_\omega^{(1)}({\bm x})={\bm t}\ee^{\ii {\bm K}_m^-\cdot{\bm x}}, \\
&{\bm K}_\mathrm{m}^\pm={\bm k}_\|\pm \Gamma_\mathrm{m}\hat{z},\\ &\Gamma_\mathrm{m}=\sqrt{q_\mathrm{m}^2-{\bm k}_\|^2},\quad q_\mathrm{m}=\frac{\omega}{c}\sqrt{\epsilon(\omega)},\\
&{\bm t}=t_p {\bm p}_\mathrm{m}^- + t_s{\bm s},\\
&{\bm p}_\mathrm{m}^\pm=\pm \frac{\Gamma_\mathrm{m}}{q_\mathrm{m}}\hat{\bm k}_\| - \frac{|{\bm k}_\||}{q_\mathrm{m}}\hat{z}, \quad {\bm s}=\hat{\bm k}_\perp,
\end{align}
where ${\bm k}_\|=(k_x,k_y)$ is the wave vector parallel to the interface, ${\bm k}_\perp=(-k_y,k_x)$,  $\hat{\bm k}_{\|(\perp)}$ is the unit vector directed to ${\bm k}_{\|(\perp)}$, and $\hat{z}$ is the unit vector directed to $z$. Coefficients $t_p$ and $t_s$ are determined via the Fresnel equation.

The second-order induced charge density $\rho_\mathrm{DC}^{(2)}$ depends solely on the $z$ coordinate, so that the electrostatic potential in the metal region is given by 
\begin{align}
&\phi_\mathrm{DC}^{(2)}(z)=a_\mathrm{m}+b_\mathrm{m}z-\frac{1}{2\epsilon_0\epsilon_{\infty}}\int_{-\infty}^0\dd z'
|z-z'|\rho_\mathrm{DC}^{(2)}(z'). \label{Eq_potential_interface}
\end{align} 
Here, the first two terms are a general solution of the homogeneous Poisson equation and the third term is a special solution of the inhomogeneous Poisson equation. 
The open-circuit condition for the photovoltage gives the Neumann-type boundary condition as 
\begin{align}
&\partial_z\phi_\mathrm{DC}^{(2)}|_{z=0}=N_z,\\
&N_z=\frac{e}{2m\omega}|{\bm t}|^2\Re\left[\frac{\ii\Gamma_\mathrm{m}}{\omega -\ii\gamma}\right],
\end{align} 
from which coefficient $b_\mathrm{m}$ is determined as 
\begin{align}
&b_\mathrm{m}=N_z+\frac{Q}{2\epsilon_0\epsilon_{\infty}},\\
&Q=\int_{-\infty}^0\dd z\rho_\mathrm{DC}^{(2)}(z). \label{Eq_bcharge_interface}
\end{align}
The explicit calculation of $Q$ indicates $Q=-\epsilon_0\epsilon_{\infty}N_z$. 
Putting $\phi_\mathrm{DC}^{(2)}(0)=0$ without the loss of generality, we obtain the following potential distribution:  
\begin{align}
\phi_\mathrm{DC}^{(2)}(z)=\frac{Q}{2\epsilon_0\epsilon_{\infty}\Im[\Gamma_\mathrm{m}]}\left(1-\ee^{2\Im[\Gamma_\mathrm{m}]z}\right). 
\end{align} 
The potential decay exponentially away from the interface and reaches to the constant value. This is the photovoltage induced in the metal.

Outside the metal, we have $\Delta\phi_\mathrm{DC}^{(2)}=0$, so that 
$\phi_\mathrm{DC}^{(2)}(z)=bz$ taking account of the continuity of the potential at the interface. 
The Maxwell's boundary condition relevant to the Gauss law results in  the induced surface charge $Q_\mathrm{s}$ that is not descried by $\rho_\mathrm{DC}^{(2)}$. It is given by 
\begin{align}
Q_\mathrm{s}=\epsilon_0\epsilon_\infty N_z-\epsilon_0b. 
\end{align}
As a natural assumption, we impose the vanishing total induced charge 
between the bulk and surface, by the second-order optical nonlinearity. Namely, 
\begin{align}
Q+Q_\mathrm{s}=0. 
\end{align}  
This gives $b=0$. Namely, the potential becomes constant.

The photocurrent in the horizontal direction becomes  
\begin{align}
&({\bm j}_\mathrm{DC}^{(2)})_\|=
\frac{e\epsilon_0\omega_\mathrm{p}^2}{2m\omega(\omega^2+\gamma^2)}
\ee^{2\Im[\Gamma_\mathrm{m}]z}\sum_{\xi,\xi'=p,s}{\bm J}_{\xi\xi'}t_\xi^*t_{\xi'},\\
&{\bm J}_{\xi\xi}=f_\xi {\bm k}_\|, \quad {\bm J}_{ps}={\bm J}_{sp}^* = -\ii \frac{\Im[\Gamma_\mathrm{m}]}{q_\mathrm{m}^*}{\bm k}_\perp, \\ &f_p=\frac{\Re[\epsilon(\omega)]}{|\epsilon(\omega)|},\quad f_s=1. \label{Eq_fxi}
\end{align}
The current can flow parallel ($\propto {\bm k}_\|$) and perpendicular ($\propto {\bm k}_\perp$) to the wave vector of the incident light, 
depending on its polarization. Particularly, if $\Re[\epsilon(\omega)]<0$, 
the current flows in the opposite direction between the P- and S-polarized incident light. Moreover, if it is circular-polarized, the perpendicular current flips its sign depending on the chirality of the left or right circular polarization.      
The spatial dependence of the current exhibits the same exponential decay as the electrostatic potential.

\section{Slab}
\begin{figure}
	\centering
	\includegraphics[width=0.45\textwidth]{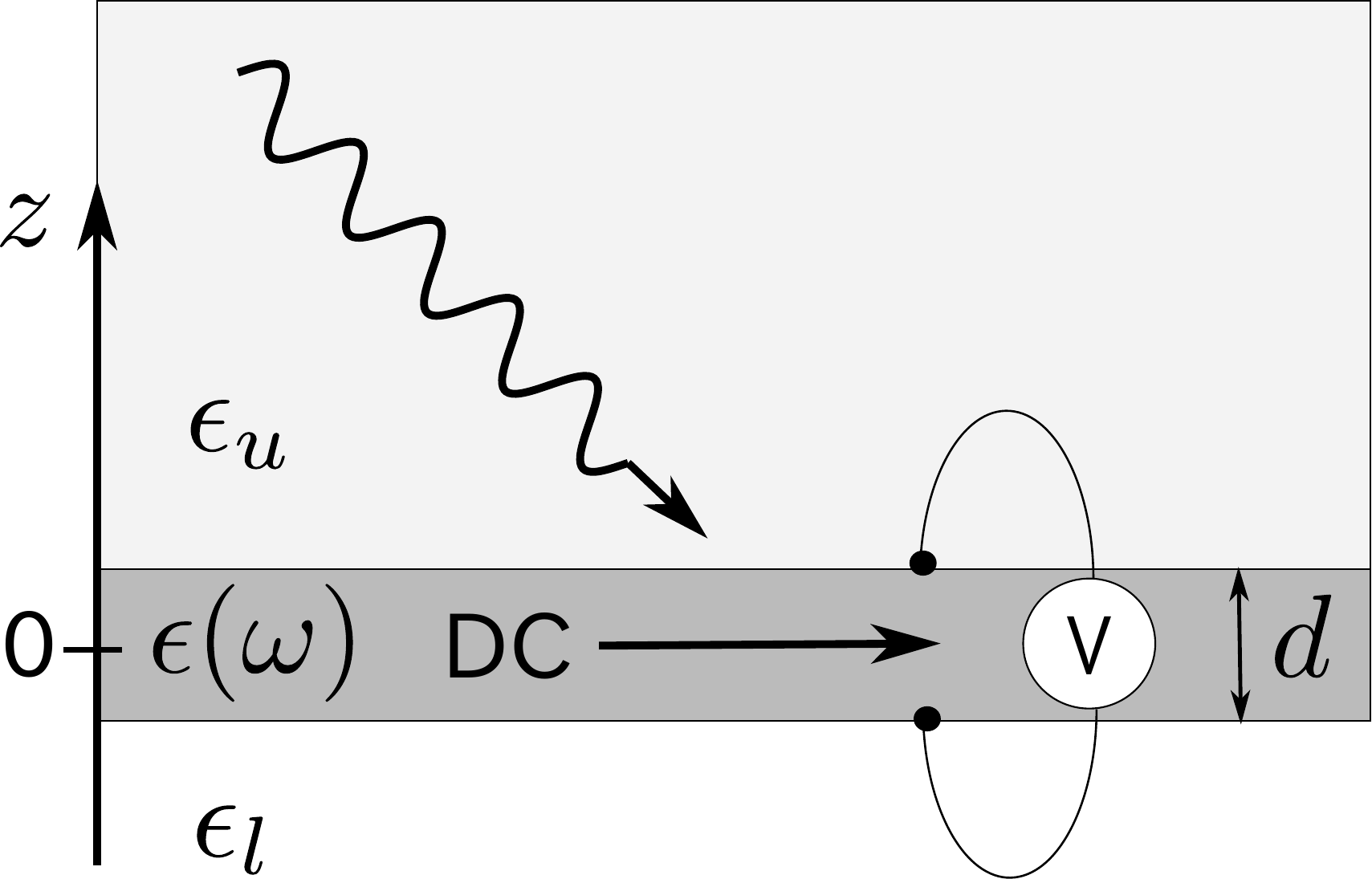}
	\caption{Schematic illustration of the slab system under study. The metallic slab with thickness $d$ is sandwiched by the semi-infinite dielectric media of permittivity $\epsilon_u$ (upper) and $\epsilon_l$ (lower).  The incident wave is coming from the upper side. The photovoltage  and photocurrent are generated in the slab.  }
	\label{fig:slabgeo}
\end{figure}

Next, let us consider a metallic slab with thickness $d$. 
A schematic illustration of the system under study is shown in Fig \ref{fig:slabgeo}.  
The incident plane wave  ${\bm E}^{(1);\mathrm{inc}}={\bm e}_0\exp(\ii{\bm k}\cdot{\bm x})$ 
is coming from the top and the resulting photovoltage between the upper and lower surfaces is probed. 
The photocurrent is generated in the horizontal direction which is assumed to be infinitely extended.

In the first order, the radiation field is solved as follows. 
By the translational invariance in plane, the electric field inside the slab is written as 
\begin{align}
&\tilde{\bm E}_\omega^{(1)}({\bm x})={\bm a}^+\ee^{\ii{\bm K}_\mathrm{m}^+\cdot{\bm x}} + {\bm a}^-\ee^{\ii{\bm K}_\mathrm{m}^-\cdot{\bm x}},\\
&{\bm a}^\pm = a_p^\pm{\bm p}_\mathrm{m}^\pm + a_s^\pm{\bm s}.
\end{align}
Outside the slab, the electric field is given by 
\begin{align}
&\tilde{\bm E}_\omega^{(1)}({\bm x})=\left\{ \begin{array}{ll}
{\bm e}_0 \ee^{\ii{\bm k}\cdot{\bm x}} + {\bm r} \ee^{\ii{\bm K}_u^+\cdot{\bm x}} & (z>\frac{d}{2}) \\
{\bm t} \ee^{\ii{\bm K}_l^-\cdot{\bm x}}  & (z<-\frac{d}{2})
\end{array}\right. \\
&{\bm e}^0=e_p^0{\bm p}_u^- + e_s^0{\bm s},\quad {\bm k}={\bm K}_u^-,\\
&{\bm r}=r_p{\bm p}_u^+ + r_s{\bm s},\quad {\bm t}=t_p{\bm p}_l^- + t_s{\bm s},\\
&{\bm p}_\eta^\pm=\pm \frac{\Gamma_\eta}{q_\eta}\hat{\bm k}_\| - \frac{|{\bm k}_\||}{q_\eta}\hat{z},\\ 
&{\bm K}_\eta^\pm={\bm k}_\|\pm \Gamma_\eta\hat{z}, \\ &\Gamma_\eta=\sqrt{q_\eta^2-{\bm k}_\|^2}, \
q_\eta=\frac{\omega}{c}\sqrt{\epsilon_\eta} \quad (\eta=u,l) 
\end{align}
where $\epsilon_{u(l)}$ is the permittivity of the upper (lower) medium. 
The Maxwell boundary condition at the slab surfaces determines all the coefficients $a_p^\pm$, $a_s^\pm$, $r_p$, $r_s$, $t_p$, and $t_s$.

By the translational invariance in plane, the induced charge density $\rho_\mathrm{DC}^{(2)}$ depends solely on the $z$ coordinate, so that 
the electrostatic potential inside the slab is given by  
\begin{align}
&\phi_\mathrm{DC}^{(2)}(z)=a_\mathrm{m}+b_\mathrm{m}z-\frac{1}{2\epsilon_0\epsilon_\infty}\int_{-\frac{d}{2}}^{\frac{d}{2}}\dd z'
|z-z'|\rho_\mathrm{DC}^{(2)}(z'). 
\end{align} 
The open-circuit condition for the photovoltage gives the Neumann-type boundary condition as 
\begin{align}
&\partial_z\phi_\mathrm{DC}^{(2)}|_{z=\pm \frac{d}{2}}=N_\pm, \\
&N_\pm=-\frac{e}{2m\omega}\Re\left[\frac{\ii\Gamma_\mathrm{m}}{\omega -\ii\gamma}\left(
|{\bm a}^+|^2 \ee^{\mp\Im[\Gamma_\mathrm{m}]d} \right.\right. \nonumber \\
& \hskip20pt -{\bm a}^{+*}\cdot{\bm a}^-\ee^{\mp\Re[\Gamma_\mathrm{m}]d} 
+{\bm a}^{-*}\cdot{\bm a}^+\ee^{\pm\Re[\Gamma_\mathrm{m}]d} \nonumber \\
& \hskip20pt \left.\left. -|{\bm a}^-|^2 \ee^{\pm\Im[\Gamma_\mathrm{m}]d}
\right)\right]. 
\end{align} 
The unknown coefficient $b_m$ is simply given by 
\begin{align}
b_m=\frac{1}{2}(N_++N_-).
\end{align}
There is a sum rule that corresponds to the Gauss law: 
\begin{align}
&N_+-N_-=-\frac{Q}{\epsilon_0\epsilon_\infty},\\
&Q=\int_{-\frac{d}{2}}^{\frac{d}{2}}\dd z\rho_\mathrm{DC}^{(2)}(z). 
\end{align}
The photovoltage between the upper and lower surfaces becomes 
\begin{align}
V&=\phi_\mathrm{DC}^{(2)}\left(\frac{d}{2}\right)-\phi_\mathrm{DC}^{(2)}\left(-\frac{d}{2}\right)\\
&=b_md + \frac{Q_z}{\epsilon_0\epsilon_\infty}, \\
Q_z&=\int_{-\frac{d}{2}}^{\frac{d}{2}}\dd zz\rho_\mathrm{DC}^{(2)}(z). 
\end{align}

Outside the metallic slab, the potential is written as 
\begin{align}
\phi_\mathrm{DC}^{(2)}(z)=\left\{\begin{array}{ll}
a_u+b_uz & (z>\frac{d}{2})\\
a_l+b_lz & (z<-\frac{d}{2})	
\end{array}\right.
\end{align} 
The Maxwell's boundary condition results in the induced surface charges as
\begin{align}
&Q_\mathrm{s}^+=\epsilon_0\epsilon_\infty N_+ -\epsilon_0\epsilon_ub_u,\\
&Q_\mathrm{s}^-=-\epsilon_0\epsilon_\infty N_- +\epsilon_0\epsilon_lb_l.
\end{align}
at the upper and lower surfaces, respectively. 
The assumption of the vanishing total induced charge becomes 
\begin{align}
0=Q+Q_\mathrm{s}^++Q_\mathrm{s}^-=-\epsilon_0(\epsilon_ub_u-\epsilon_lb_l). 
\end{align}
We can put $b_u=b_l=0$, otherwise a nonzero electric field emerges 
with the same orientation in the upper and lower media, which is not physically acceptable. 
Also, we can put $a_\mathrm{m}=0$ without the loss of generality. 
The continuity of the potential itself gives 
\begin{align}
&a_u=N_+\frac{d}{2}+\frac{Q_z}{2\epsilon_0\epsilon_\infty},\\
&a_l=-N_-\frac{d}{2}-\frac{Q_z}{2\epsilon_0\epsilon_\infty}.
\end{align}
In this way, the potential profile is fully determined.

The photocurrent in the horizontal direction is given by  
\begin{align}
&{\bm J}_\|=W\int_{-\frac{d}{2}}^{\frac{d}{2}} \dd z ({\bm i}_\mathrm{DC}^{(2)})_\| \\
&\hskip10pt =\frac{e\epsilon_0\omega_\mathrm{p}^2W}{2m\omega(\omega^2+\gamma^2)}\sum_{\zeta,\zeta'=p,s}\sum_{\sigma,\sigma'=+,-}{\bm J}_{\zeta\zeta'}^{\sigma\sigma'}a_\zeta^{\sigma*}a_{\zeta'}^{\sigma'},\\
&{\bm J}_{\zeta\zeta}^{++}={\bm J}_{\zeta\zeta}^{--}=f_\zeta \frac{\sinh(\Im[\Gamma_\mathrm{m}]d)}{\Im[\Gamma_\mathrm{m}]}{\bm k}_\|,\\ 
&{\bm J}_{\zeta\zeta}^{+-}={\bm J}_{\zeta\zeta}^{-+}=f_\zeta \frac{\sin(\Re[\Gamma_\mathrm{m}]d)}{\Re[\Gamma_\mathrm{m}]}{\bm k}_\|,\\ 
&{\bm J}_{ps}^{++}=-{\bm J}_{ps}^{--}={\bm J}_{sp}^{++*}=-{\bm J}_{sp}^{--*}
=\ii \frac{\sinh(\Im[\Gamma_\mathrm{m}]d)}{q_\mathrm{m}^*}{\bm k}_\perp,\\
&{\bm J}_{ps}^{+-}=-{\bm J}_{ps}^{-+}=-{\bm J}_{sp}^{+-*}={\bm J}_{sp}^{-+*}=-\frac{\sin(\Re[\Gamma_\mathrm{m}]d)}{q_\mathrm{m}^*}{\bm k}_\perp,
\end{align}
where $W$ is the relevant sample width in plane. 
The mixing between the P- and S-polarization gives rise to the term proportional to ${\bm k}_\perp$. Namely, the photocurrent is induced in the transverse direction to the incident light.  For circular-polarized light, the transverse photocurrent changes its sign by switching the chirality of the circular polarization.

The above photocurrent is a short-circuit current assuming the infinite extent of the slab in plane. By the translational invariance in plane, this implies the vanishing photovoltage in the horizontal direction.  
If the slab is finite, the translational invariance is broken, so that the photovoltage becomes nonzero under the open-circuit condition. 
The resulting  photovoltage  can be roughly estimated by assuming vanishing current density in horizontal direction, in the sample bulk. 
For instance, in the $x$ direction, we have the photovoltage $V_x$ as  
\begin{align}
V_x\sim \frac{m\gamma}{e\rho_0}\frac{L}{Wd}J_x,
\end{align}
where $L$ is the length of the film in the $x$ direction, and we assume the average over the slab in the $z$ direction. 
We should note that the above result is of the rough estimate, and that the correct value of the photovoltage in the horizontal direction is obtained by solving the electrostatic potential problem under the Neumann boundary condition. The resulting value must be correlated with the photovoltage in the $z$ direction, through a (probably) complex electrostatic-potential distribution.  

In what follows, we assume that the system is 
made of gold with the following Drude fitting parameters:  $\epsilon_\infty=9.84,\;\hbar\omega_\mathrm{p}=9.01\;[\mathrm{eV}],\; \hbar\gamma=0.072\; [\mathrm{eV}]$. 
The fitting is reasonable from infrared to visible frequency ranges below the interband transition threshold.

\begin{figure}
	\centering
\includegraphics[width=0.45\textwidth]{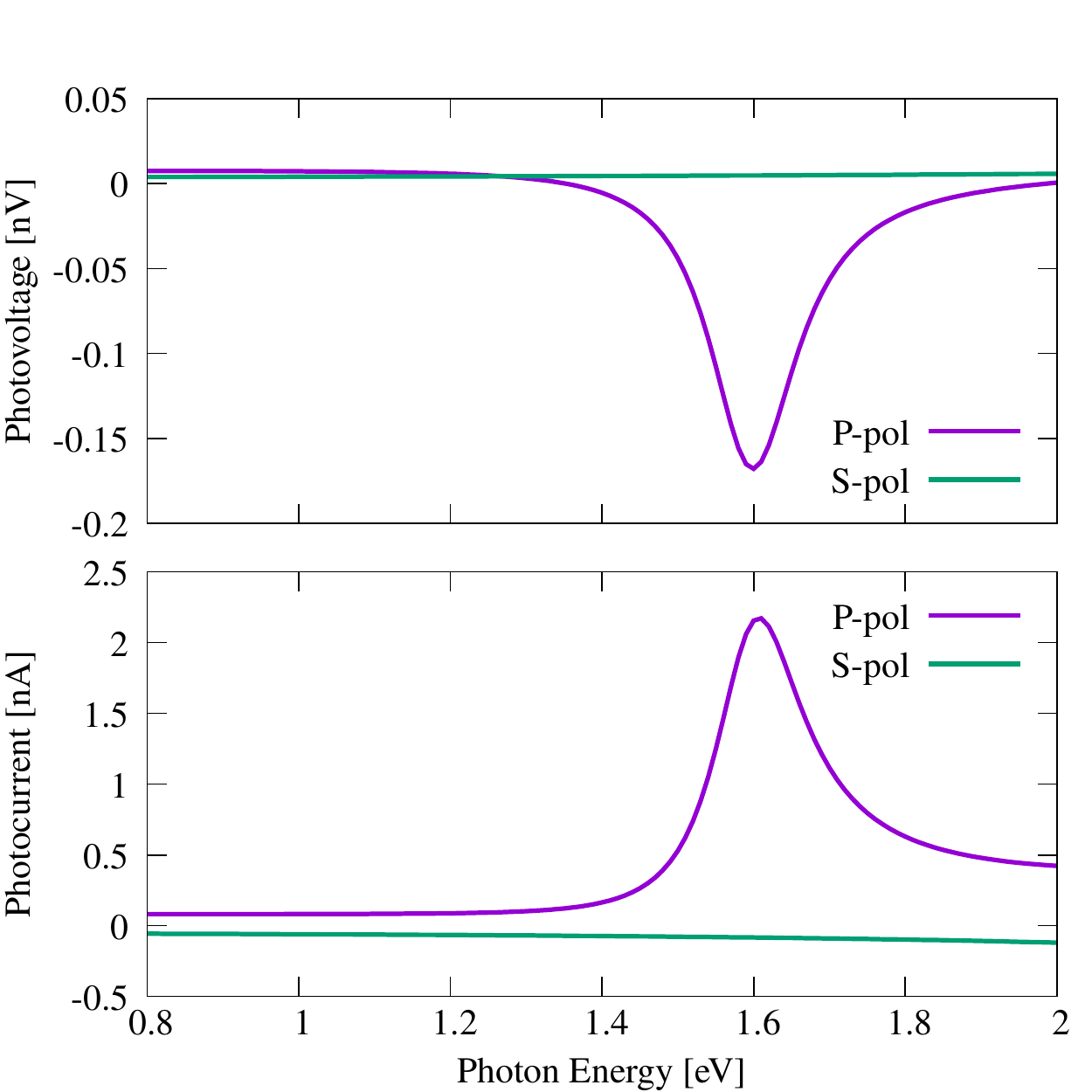}
	\caption{Photovoltage $V$ and photocurrent $J_\|$ generated in the Au slab of thickness $d$=50 [nm] deposited to a glass of permittivity $\epsilon_u=2.1$. The incident  plane-wave light is coming from the glass side with the incident angle $\theta_\mathrm{inc}=45^\circ$ and  the vacuum flux density $F\equiv\epsilon_0c|{\bm e}^0|^2/2$=500 [W/cm${}^2$].  The sample width is assumed to be $W$=100 [$\mu$m] perpendicular to the incident plane. The term "P(S)-pol" refers to the P(S) polarization of the incident light, in which the electric field polarization is parallel (perpenducular) to the incident plane. 
		The photovoltage and photocurrent are linear in the flux density, so that $V$ and $J_\|$ at other $F$ are equal to the graph values multiplied by the ratio of $F$ and 500 [W/cm${}^2$].}
	\label{fig:au50nmpde}
\end{figure}

Figure \ref{fig:au50nmpde} shows the spectrum of the photovoltage between the upper and lower surfaces of a gold slab deposited to 
a semi-infinite glass of $\epsilon_u=2.1$.  The lower medium has $\epsilon_l=1$. 
The attenuated total-internal reflection (ATR) setting with the Kretschmann configuration is assumed for the surface plasmon excitation \cite{kretschmann1968radiative}.
It is remarkable that the photovoltage and photocurrent for the P-polarized incident light 
exhibit a resonance at $\hbar\omega$=1.6 [eV], which corresponds to the surface plasmon of the gold-air interface. In contrast, the spectra for the S-polarized light are plain.   
  
\begin{figure}
	\centering
\includegraphics[width=0.45\textwidth]{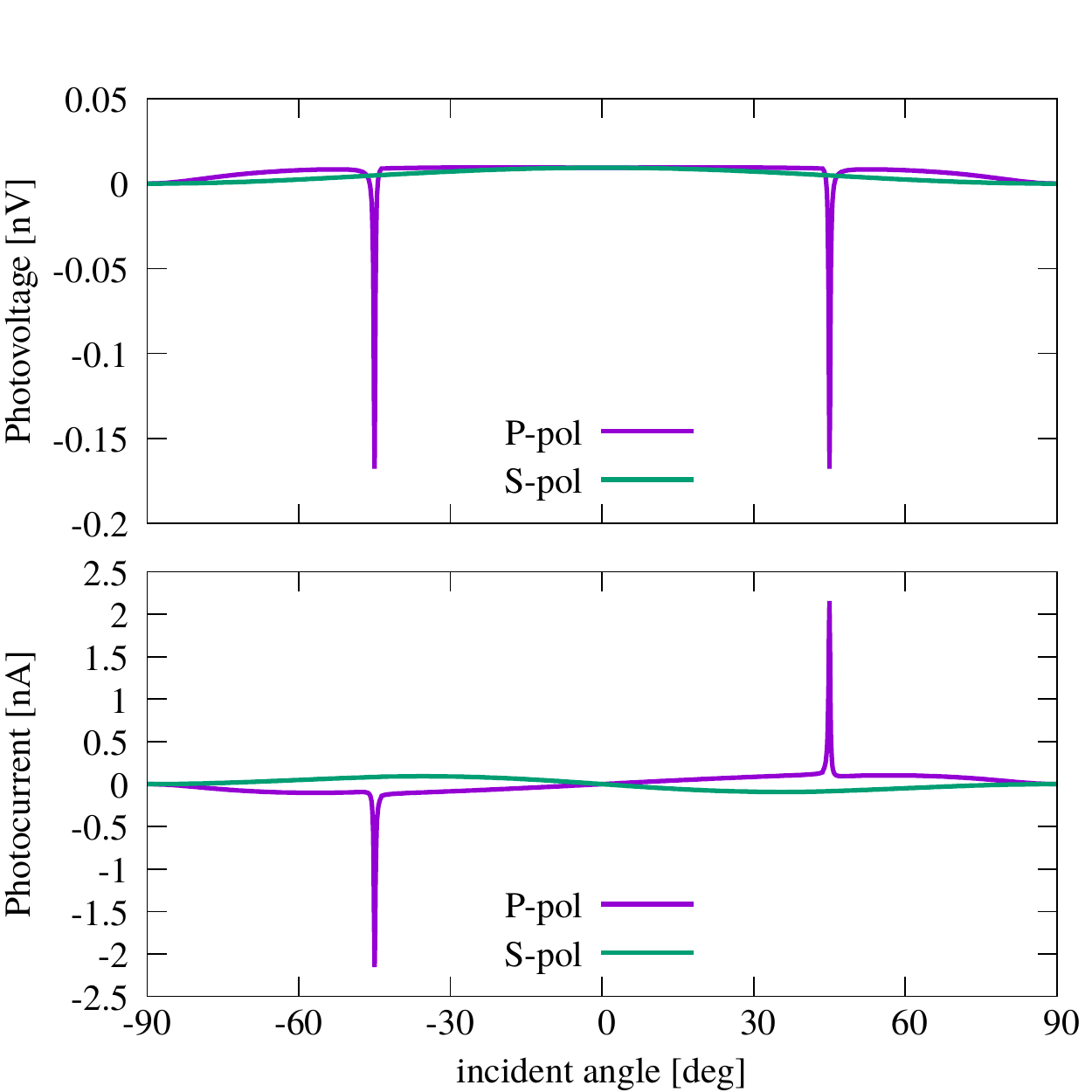}
	\caption{Incident angle dependence of photovoltage and photocurrent in the Au slab of the Kretschmann configuration.   
       The frequency of the incident light is fixed to $\hbar\omega$=1.6 [eV]. The other parameters are the same as in  Fig. \ref{fig:au50nmpde}, except for the incident angle.   }
	\label{fig:au50nmpdeome14thinc}
\end{figure}

Figure \ref{fig:au50nmpdeome14thinc} shows the incident-angle dependence of the photovoltage and photocurrent at the resonant frequency.  
The photovoltage and photocurrent exhibit the sharp peak and dip at the incident angle of $\pm 45^\circ$ for the P-polarization. The off-resonant magnitudes of the photovoltage and photocurrent are of the same order between the P and S polarizations. The signs of the photocurrent are opposite between P and S, owing to the factor $f_\xi$ of Eq. (\ref{Eq_fxi}).   
As expected by the inversion symmetry in plane, the photovoltage is symmetric with respect to the incident angle, whereas the photocurrent is anti-symmetric.

\begin{figure}
	\centering
\includegraphics[width=0.45\textwidth]{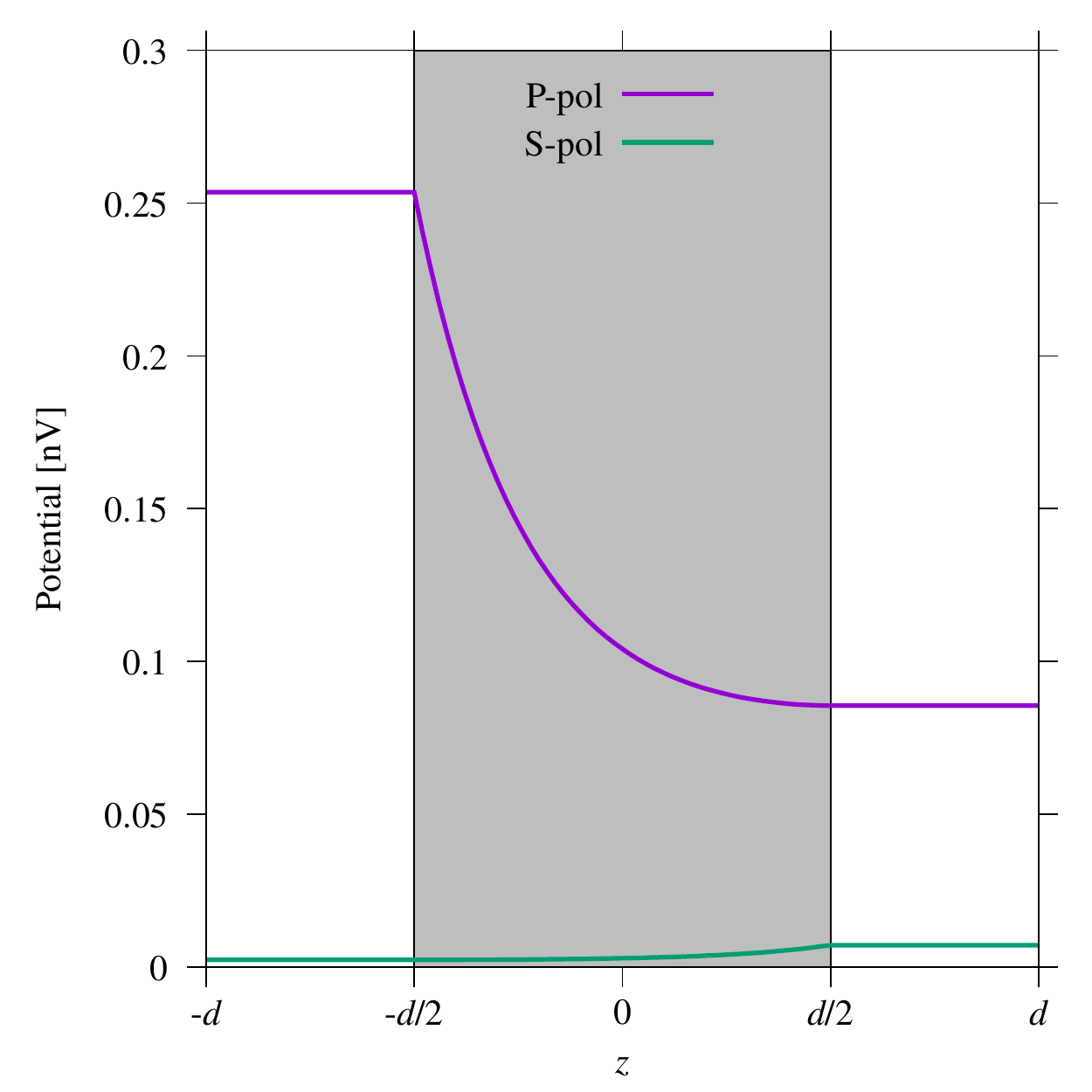}
	\caption{Electrostatic-potential distribution in the Au slab. The incident plane-wave light is coming from the positive $z$ region with photon energy of $\hbar\omega$=1.6 [eV] and the incident angle of $45^\circ$.  The other parameters are the same as in Fig. \ref{fig:au50nmpde}.} 	
	\label{fig:au50nmpdeome14}
\end{figure}

Figure \ref{fig:au50nmpdeome14} shows the electrostatic-potential distribution inside and outside the slab. 
For the P polarization, the potential grows exponentially inside the slab toward the bottom surface ($z=-d/2$), showing the localization nature of the surface plasmon at the gold-air interface. 
Outside the slab, the potential is constant, because the total induced charge cancels between the bulk and surface. The discontinuity of the potential slope indicates the nonzero induced surface charge.    
The potential difference through the Au slab corresponds to the resonance dip 
of the photovoltage in Figs. \ref{fig:au50nmpde} and \ref{fig:au50nmpdeome14thinc}. 

We also note that the local photocurrent $j_\mathrm{DC}^{(2)}$ in the $z$ direction completely vanishes by the translational invariance in plane and the open-circuit condition at the surfaces, implying the cancellation between the Ohmic and EMF currents.

\section{Cylinder}
\begin{figure}
	\centering
		\includegraphics[width=0.45\textwidth]{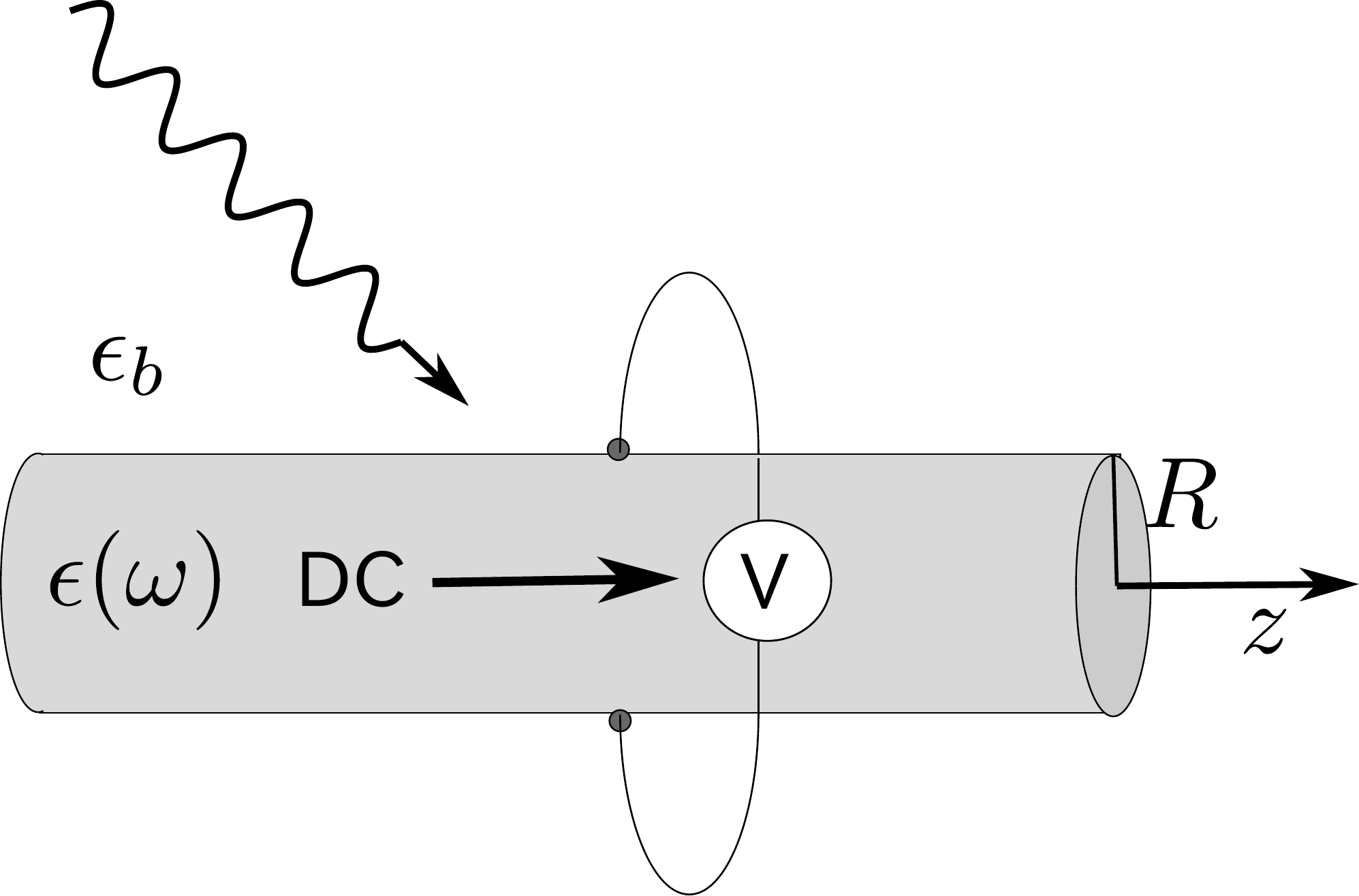}
	\caption{A schematic illustration of the metallic cylindrical system under study. It has infinite length in the $z$ direction and radius $R$. The medium outside the cylinder has  permittivity $\epsilon_b$.   
	    The incident light is coming from the outside, and the photovoltage is measured at the cylinder surface. 
		}
	\label{fig:cylgeo}
\end{figure}

Next, let us consider a metallic cylinder with infinite length and radius $R$. 
A schematic illustration of the system under study is shown in Fig. \ref{fig:cylgeo}. 

In the linear response, the electric field inside the cylinder is given by  
\begin{align}
&\tilde{\bm E}_\omega^{(1)}({\bm x})=\sum_{nn'}\sum_{\zeta=\mathrm{M,N}} J_n(\lambda_\mathrm{m}r)\ee^{\ii n\varphi}\ee^{\ii k_zz}{\bm P}_{nn'}^\zeta\psi_{n'}^{\zeta<},\\
&\lambda_\mathrm{m}=\sqrt{q_\mathrm{m}^2-k_z^2},
\end{align}
where $k_z$ is the wave number along the cylindrical axis, $(r,\varphi,z)$ is the cylindrical coordinate, $J_n$ is the Bessel function of integer order $n$, ${\bm P}^\zeta$ is the vector-cylindrical-wave matrix \cite{Ohtaka:U:A::57:p2550-2568:1998}, and $\psi_n^{\zeta <} \; (\zeta=\mathrm{M,N})$ are the vector-cylindrical-wave-expansion  coefficients.  
Outside the cylinder, the electric field is given by 
\begin{align}
&\tilde{\bm E}_\omega^{(1)}({\bm x})={\bm e}^0 \ee^{\ii {\bm k}\cdot{\bm x}} \nonumber \\
&\hskip35pt +\sum_{nn'}\sum_{\zeta=\mathrm{M,N}} H_n(\lambda_br)\ee^{\ii n\varphi}\ee^{\ii k_zz}{\bm P}_{nn'}^\zeta\psi_{n'}^{\zeta >},\\
&{\bm k}=q_b(\sin\theta_\mathrm{inc}\cos\varphi_\mathrm{inc},\sin\theta_\mathrm{inc}\sin\varphi_\mathrm{inc},\cos\theta_\mathrm{inc}),\\
&\lambda_b=\sqrt{q_b^2-k_z^2},\quad q_b=\frac{\omega}{c}\sqrt{\epsilon_b}.
\end{align}
where $H_n$ is the Hankel function of the first kind and integer order $n$. The coefficients 
$\psi_n^{\zeta >(<)}$ are determined via the S-matrix of  the cylinder \cite{bohren2008absorption}.

By the translational invariance in the $z$ direction, the second-order charge density $\rho_\mathrm{DC}^{(2)}$ is independent of the $z$ coordinate, so that the electrostatic potential is a function of ${\bm x}_\|=(x,y)$ and is given by  
\begin{align}
&\phi_\mathrm{DC}^{(2)}({\bm x}_\|)= \sum_{n}a_nr^{|n|}\ee^{\ii n \varphi} \nonumber \\
&\hskip30pt  -\frac{1}{2\pi\epsilon_0\epsilon_\infty}\int_{r'\le R} d^2{\bm x}'_\|\log|{\bm x}_\|-{\bm x}'_\|| \rho_\mathrm{DC}^{(2)}({\bm x}'_\|), 
\end{align}
with unknown coefficients $a_n$, inside the cylinder. 
The boundary condition for the electrostatic potential at the cylinder surface is of the Neumann type: 
\begin{align}
\partial_r\phi_\mathrm{DC}^{(2)}|_{r=R}&=\frac{\gamma}{\epsilon_0\omega_\mathrm{p}^2}({\bm i}_\mathrm{DC}^{(2)})_r|_{r=R}\nonumber \\
&=\sum_nN_n\ee^{\ii n\varphi}, 
\end{align}
which determines the coefficients $a_n$ as 
\begin{align}
&|n|a_nR^{|n|-1}=N_n+\frac{Q_0}{2\pi\epsilon_0\epsilon_\infty R}\delta_{n,0}\nonumber \\
&\hskip77pt +\frac{Q_n}{4\pi\epsilon_0\epsilon_\infty R^{|n|+1}}(1-\delta_{n,0}), \\
&Q_n=\int_{r<R}\dd^2{\bm x}_\| r^{|n|}\ee^{-\ii n\varphi}\rho_\mathrm{DC}^{(2)}({\bm x}_\|). 
\end{align}

The electrostatic potential outside the cylinder  is simply given by 
\begin{align}
&\phi_\mathrm{DC}^{(2)}({\bm x}_\|)= \sum_{n}b_nr^{-|n|}\ee^{\ii n \varphi}, \\
&b_n=a_nR^{2|n|}+\frac{Q_n}{4\pi\epsilon_0\epsilon_\infty|n|} (1-\delta_{n,0}), 
\end{align}
where coefficients $b_n$ are determined by the continuity of the potential itself at the cylinder surface.

Imposing the Maxwell boundary condition results in the induced surface-charge density $\sigma(\varphi)$ as 
\begin{align}
&\sigma(\varphi)=\sum_n\sigma_n\ee^{\ii n\varphi},\\
&\sigma_0=-\frac{Q_0}{2\pi R},\label{Eq_Qs_cylinder}\\
&\sigma_n=\epsilon_0(\epsilon_\infty+\epsilon_b)N_n+\frac{\epsilon_bQ_n}{2\pi \epsilon_\infty}R^{-|n|-1}\quad (n\ne 0). 
\end{align}  
Equation (\ref{Eq_Qs_cylinder}) implies that the total induced charge cancels between the bulk ($Q_0$) and surface ($Q_\mathrm{s}=2\pi R\sigma_0$).

If we measure the photovoltage $V$ between ${\bm x}_\|=\pm R\hat{x}$, 
we have 
\begin{align}
V&=\phi_\mathrm{DC}^{(2)}(R\hat{x})-\phi_\mathrm{DC}^{(2)}(-R\hat{x})\nonumber \\
&=\sum_{n=\mathrm{odd}}\left( 2a_nR^{|n|}+\frac{Q_n}{2\pi\epsilon_0\epsilon_\infty |n|R^{|n|}}\right). 
\end{align}

The photocurrent is also induced in the $z$ direction. 
We evaluate it simply by assuming a uniformity of the electrostatic potential along the $z$ direction. The result is given by 
\begin{align}
&J_z=\int_{r<R}\dd^2{\bm x}_\|({\bm i}_\mathrm{DC}^{(2)})_z \nonumber\\
&\hskip10pt =\frac{e\epsilon_0\omega_\mathrm{p}^2}{2m\omega(\omega^2+\gamma^2)}\sum_n\sum_{\zeta,\zeta'(=\mathrm{M,N})} J_{n}^{\zeta\zeta'}\psi_n^{\zeta <*}\psi_n^{\zeta'<},\\
&J_{n}^\mathrm{MM}= \frac{k_z}{(\lambda_\mathrm{m}^*)^2-\lambda_\mathrm{m}^2}\frac{1}{2}(f_{l-1}+f_{l+1}),\\
&J_{n}^\mathrm{MN}=\frac{k_z}{(\lambda_\mathrm{m}^*)^2-\lambda_\mathrm{m}^2}\frac{k_z}{2q_\mathrm{m}}(f_{l-1}-f_{l+1})\nonumber \\
&\hskip20pt +R|J_n(\lambda_\mathrm{m})|^2 \frac{l\lambda_\mathrm{m}}{2q_\mathrm{m}\lambda_\mathrm{m}^*R},\\
&J_{n}^\mathrm{NM}=\frac{k_z}{(\lambda_\mathrm{m}^*)^2-\lambda_\mathrm{m}^2}\frac{k_z}{2q_\mathrm{m}^*}(f_{l-1}-f_{l+1})\nonumber \\
&\hskip20pt +R|J_n(\lambda_\mathrm{m})|^2 \frac{l\lambda_\mathrm{m}^*}{2q_\mathrm{m}^*}\lambda_\mathrm{m}R,\\
&J_{n}^\mathrm{NN}=\frac{k_z}{(\lambda_\mathrm{m}^*)^2-\lambda_\mathrm{m}^2}\left(\frac{k_z^2}{2|q_\mathrm{m}|^2}(f_{l-1}+f_{l+1})+ |\frac{\lambda_\mathrm{m}}{q_\mathrm{m}}|^2 f_l\right)\nonumber\\
&\hskip20pt +2R\frac{k_z}{|q_\mathrm{m}|^2}\Re[\lambda_\mathrm{m}^*J_l(\lambda_\mathrm{m}R)J'_l(\lambda_\mathrm{m}^*R)],\\
&f_l=\lambda_\mathrm{m}RJ_l(\lambda_\mathrm{m}R)J'_l(\lambda_\mathrm{m}^*R)-\lambda_\mathrm{m}^*RJ_l(\lambda_\mathrm{m}R)J'_l(\lambda_\mathrm{m}^*R).
\end{align}

If the cylinder has finite length $L$ in the $z$ direction, the photovoltage $V_z$ in the $z$ direction is roughly estimated by 
\begin{align}
V_z\sim \frac{m\gamma}{e\rho_0}\frac{L}{\pi R^2}J_z.  
\end{align} 
However, a correct evaluation of the photovoltage is only through solving the Poisson equation for the finite-size cylinder with the Neumann boundary condition.

\begin{figure}
	\centering
	\includegraphics[width=0.45\textwidth]{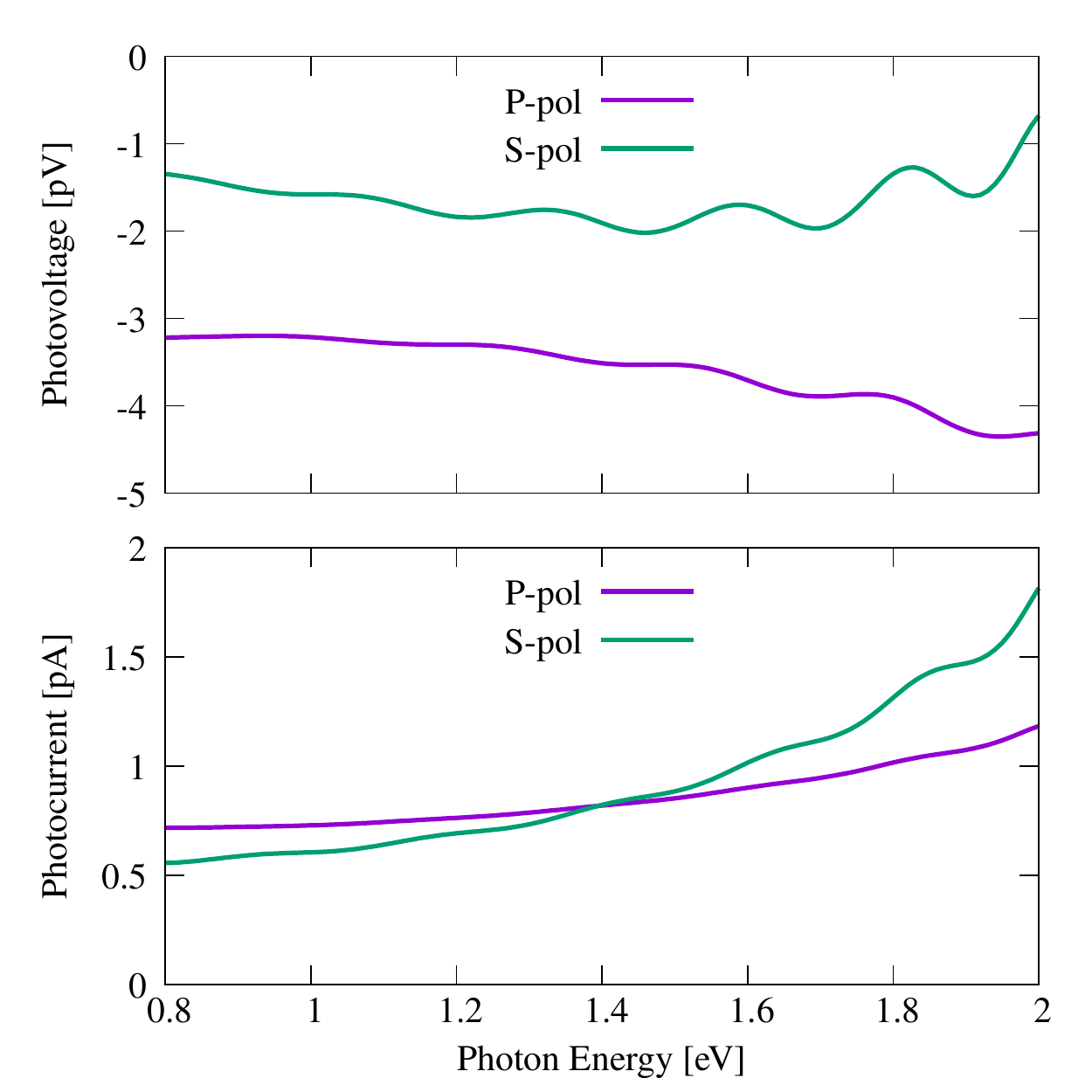}
	\caption{Photovoltage and photocurrent spectra in an isolated Au cylinder as a function of frequency of the incident light. The cylinder radius is 1 [$\mu$m]. 
	The incident angle is fixed as $(\theta_\mathrm{inc},\varphi_\mathrm{inc})=(45^\circ,30^\circ)$. The flux density of the incident light is $\epsilon_0c|{\bm e}^0|^2/2$=500 [W/cm${}^2$]. The medium outside the cylinder is air ($\epsilon_b=1$).  The P- and S-polarization are referred to the electric field polarization parallel and perpendicular to the incident plane, respectively, which is formed by the incident wave vector and the cylindrical axis. }
	\label{fig:aur1micrt45p30psor}
\end{figure}

Figure \ref{fig:aur1micrt45p30psor} shows the photovoltage $V$ and photocurrent $J_z$ in an isolated Au cylinder as a function of frequency of the incident light. 
The spectra oscillate in frequency. This oscillation is due to a weak interference relevant to the S-matrix of the cylinder. 
We should note that the absolute values of the photovoltage and photocurrent are much lower than in the slab case presented in Sec. IV. These low signals come from the fact that 
the propagating surface plasmons in the cylinder cannot be excited in the present setting without the ATR.  
However, their trace still exist as an excitation of leaky surface plasmons, or in other words, the Mie resonances.

\begin{figure}
	\centering 
	\includegraphics[width=0.45\textwidth]{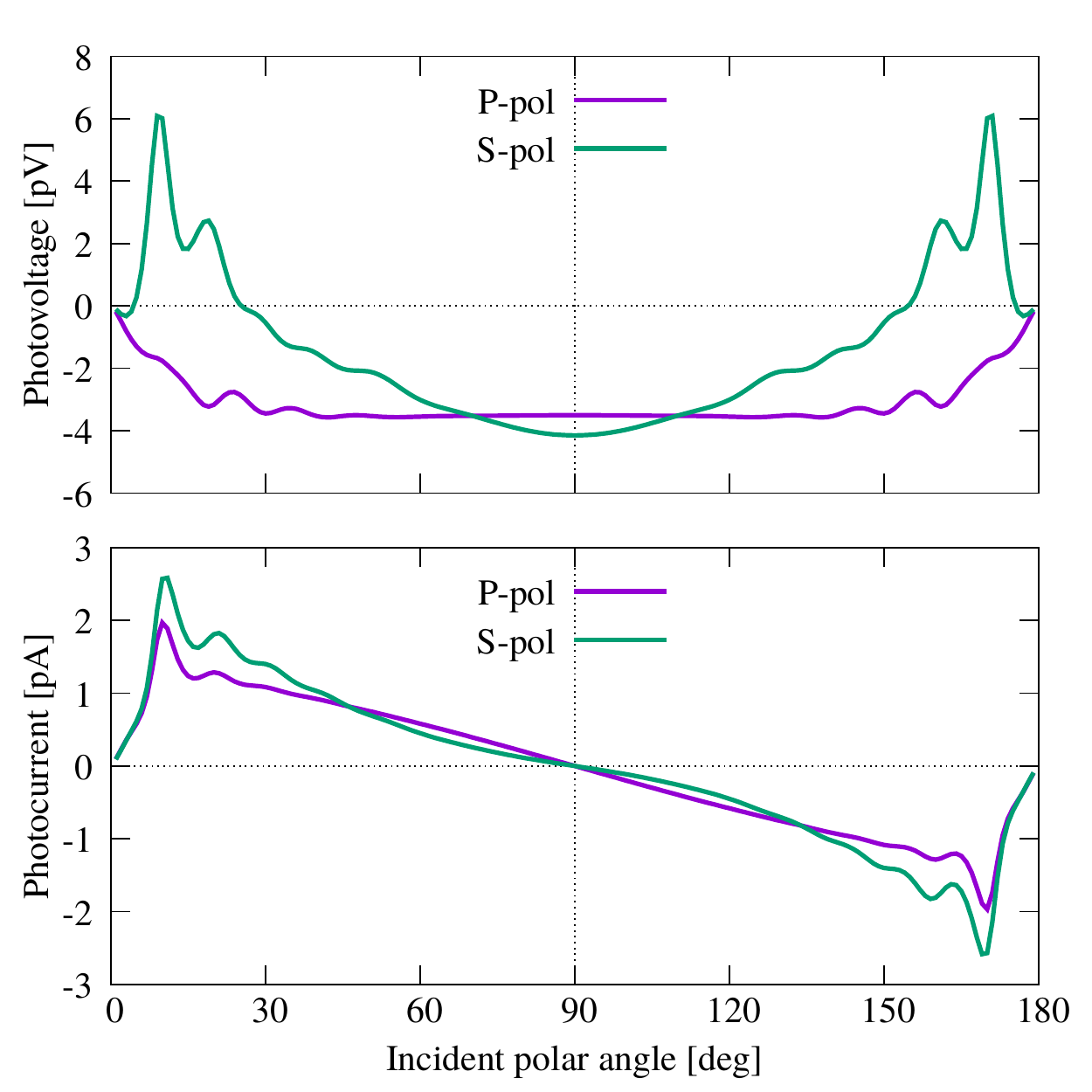}
	\caption{Incident polar-angle dependence of the photovoltage and photocurrent in the Au cylinder at $\hbar\omega$=1.46 [eV]. The incident azimuthal angle is fixed to $\varphi_\mathrm{inc}=30^\circ$. The other parameters are the same as in Fig. \ref{fig:aur1micrt45p30psor}. }
	\label{fig:aur1micrpw146psor}
\end{figure}

Figure \ref{fig:aur1micrpw146psor} shows the incident polar-angle  dependence of the photovoltage and photocurrent at $\hbar\omega$=1.46 [eV], where a  photovoltage dip is found for the S-polarization. 
We can see marked peaks at $\theta_\mathrm{inc}\sim 10^\circ, 170^\circ$ particularly for the S-polarization. These are the Mie resonance signal that can be excited without using the ATR. 
We can find the corresponding sharp absorption peaks there in the linear response (not shown).  
We also note that the photovoltage (photocurrent) exhibits a symmetric (anti-symmetric) dependence on $\theta_\mathrm{inc}$ with respect to $\theta_\mathrm{inc}=90^\circ$, as expected by the inversion symmetry in the $z$ direction.

\begin{figure}
	\centering 
\includegraphics[width=0.45\textwidth]{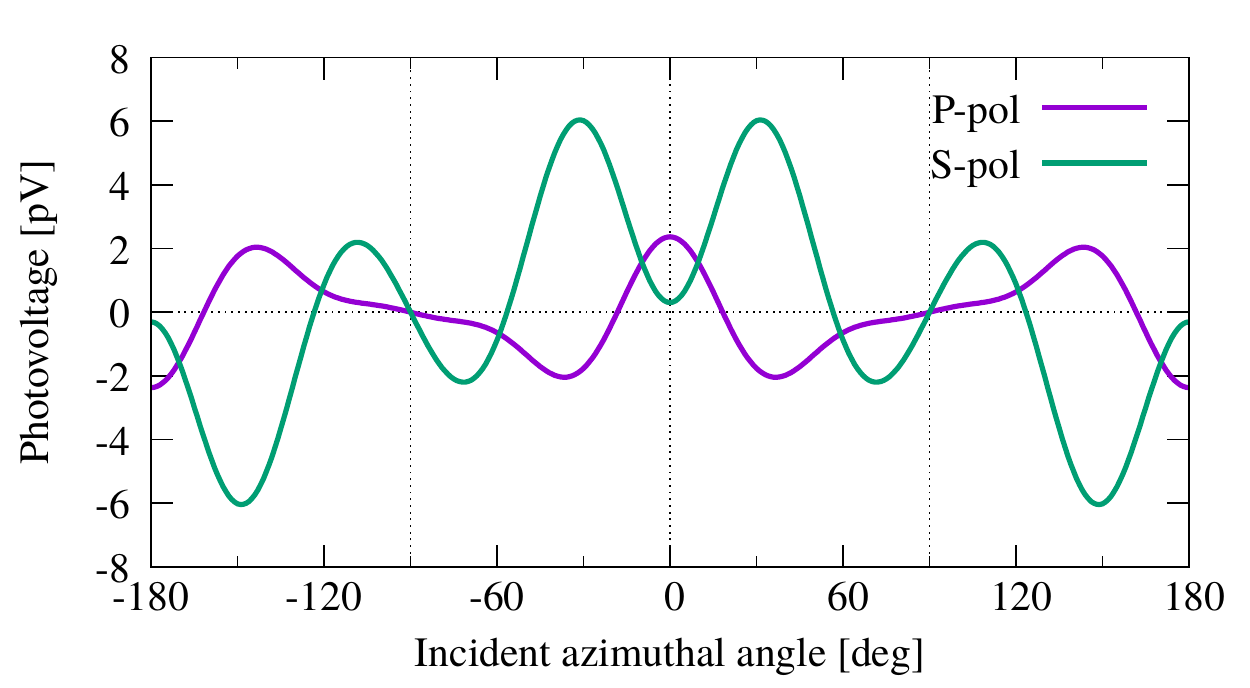}
	\caption{Incident azimuthal-angle dependence of the photovoltage in the Au cylinder at $\hbar\omega$=1.46 [eV]. The incident polar angle is fixed to $\theta_\mathrm{inc}=10^\circ$. The other parameters  are the same as in Fig. \ref{fig:aur1micrt45p30psor}.  }
	\label{fig:aur1micrtw146psor}
\end{figure}

Fixing the polar angle at a peak position ($\theta_\mathrm{inc}=10^\circ$) of Fig. \ref{fig:aur1micrpw146psor},  we show the azimuthal-angle dependence of the photovoltage in Fig. \ref{fig:aur1micrtw146psor}.
We can see that the photovoltage is symmetric with respect to $\varphi=0^\circ,190^\circ$, and is anti-symmetric with respect to $\varphi=\pm 90^\circ$, so that the photovoltage vanishes there. 
The photocurrent does not change by the incident azimuthal angle by the rotational symmetry of the system.

\begin{figure}
	\centering
\includegraphics[width=0.45\textwidth]{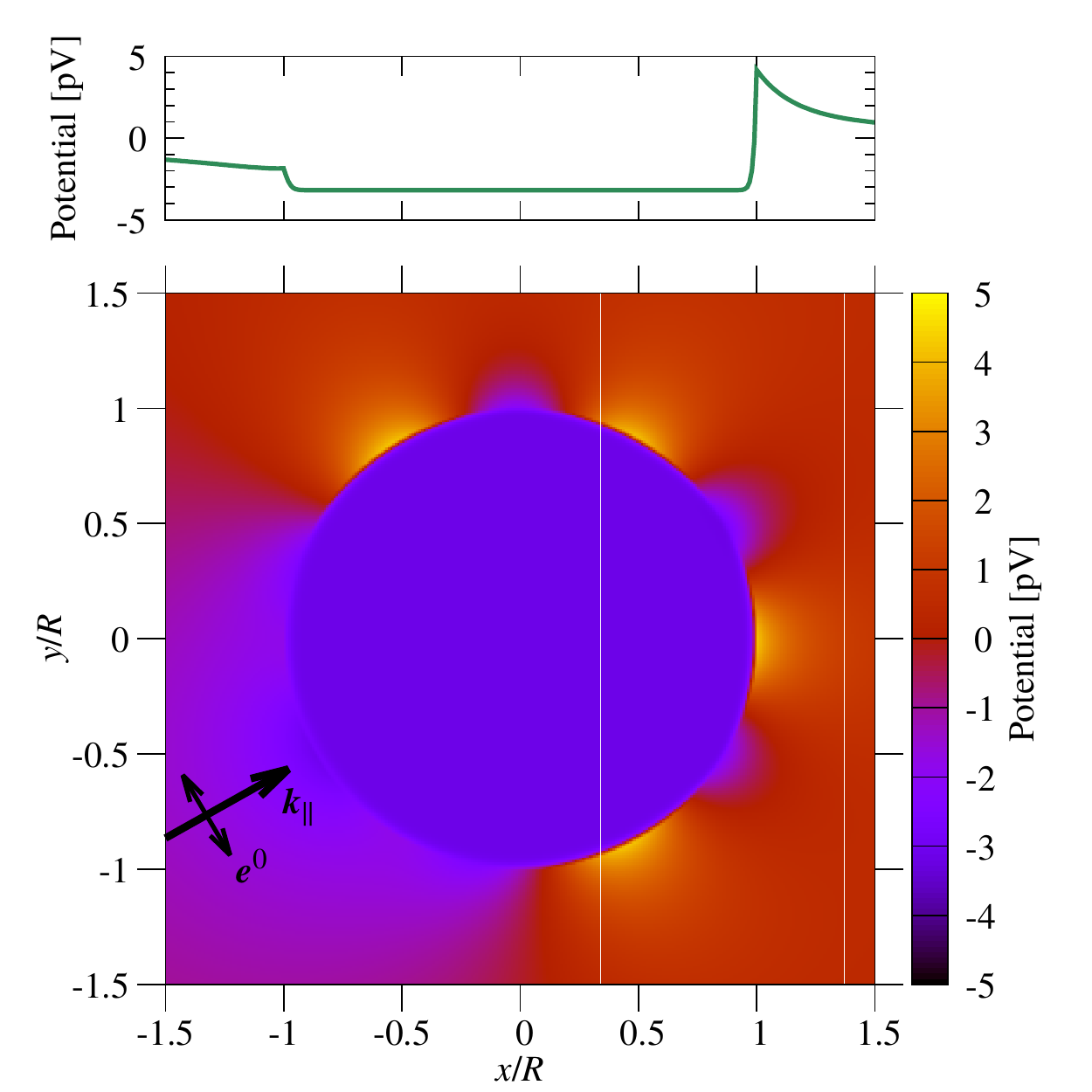}
	\caption{Electrostatic potential distribution around the Au cylinder, for the incident S-polarized light of $\hbar\omega$=1.46 [eV]. The incident angle is $(\theta_\mathrm{inc},\varphi_\mathrm{inc})=(10^\circ,30^\circ)$. The other parameters are the same as in Fig. \ref{fig:aur1micrt45p30psor}. The potential is taken to be zero at $r\to\infty$. The upper panel is the cross-sectional plot at $y=0$.  }
	\label{fig:aur1micrt10p30w146sV}
\end{figure}

Figure \ref{fig:aur1micrt10p30w146sV} shows the potential distribution at the Mie resonance.  
It has equal-spacing (of angle $60^\circ$) hot spots around the cylinder, indicating the Mie resonance of angular momentum $n=\pm 3$.  
Inside the cylinder, the potential becomes  nearly a constant. 
We can see the symmetric distribution of the potential with respect to the wave vector of the incident light oriented to the direction of $\varphi=30^\circ$.

\section{Sphere}

\begin{figure}
	\centering
\includegraphics[width=0.35\textwidth]{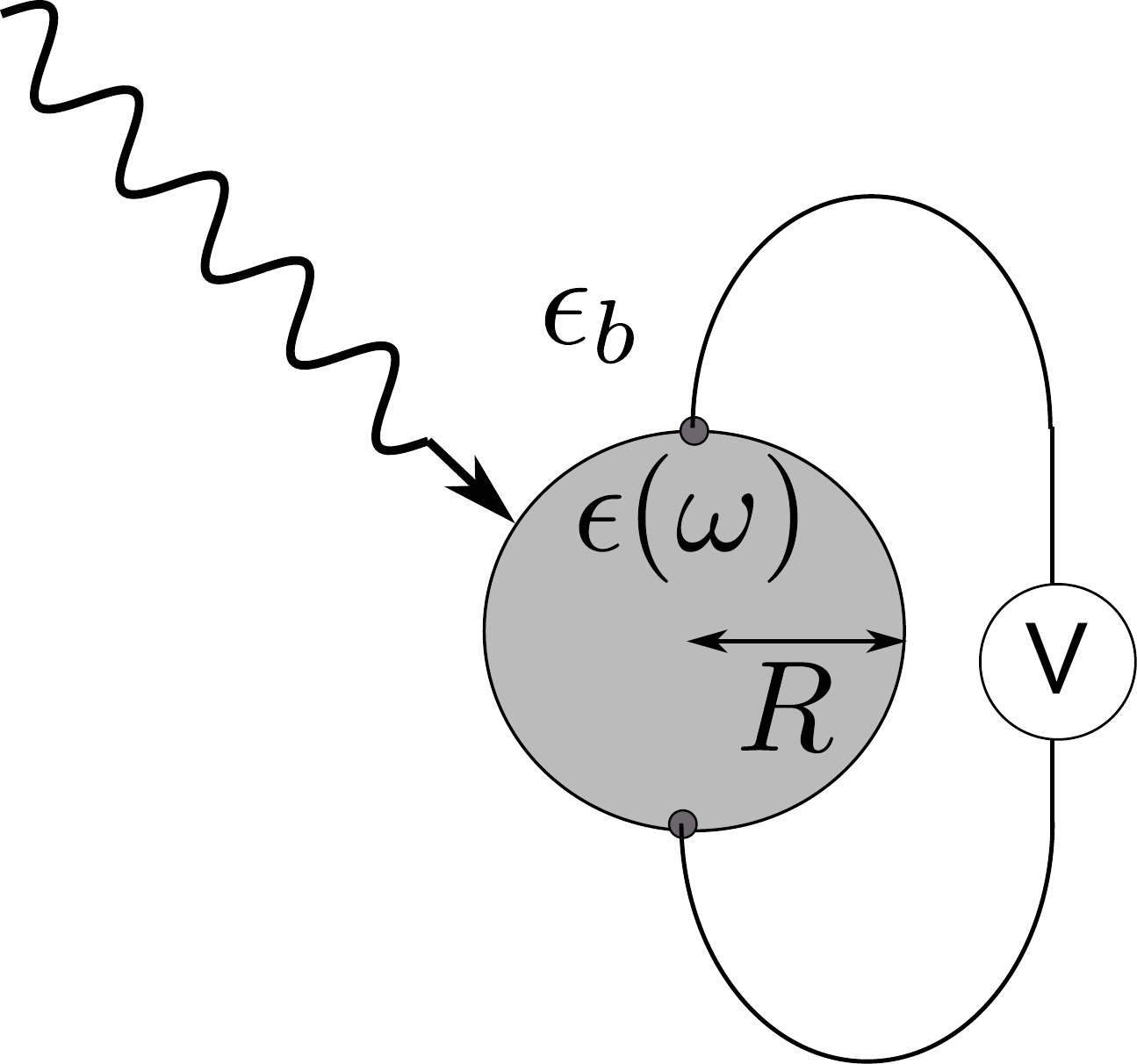}
	\caption{Schematic illustration of the spherical system under study. A metallic sphere with radius $R$ is illuminated by a plane-wave light and the photovoltage between the north and south poles is probed. The medium outside the sphere has permittivity $\epsilon_b$.  }
	\label{fig:sphgeo}
\end{figure}

Finally, let us consider a metallic sphere. 
A schematic illustration of the system under study is shown in Fig. \ref{fig:sphgeo}. 

In the linear response, the electric field inside the sphere is written as 
\begin{align}
&\tilde{\bm E}_\omega^{(1)}({\bm x})=\sum_{LL'}\sum_{\zeta=\mathrm{M,N}}j_l(q_\mathrm{m}r)Y_L(\Omega){\bm P}_{LL}^\zeta \psi_{L'}^{\zeta <},
\end{align} 
where $(r,\Omega)$ is the spherical coordinate, $L=(l,m)$ the angular momentum indices satisfying $|m|\le l$, $j_l$ is the spherical Bessel function of integer order $l$, $Y_L$ is the spherical harmonics, ${\bm P}^\zeta$ is the vector-spherical-wave-expansion matrix \cite{OHTAKA::19:p5057-5067:1979}, and $\psi_{L}^{\zeta <}\; (\zeta=\mathrm{M,N})$ are the vector-spherical-wave-expansion coefficients.       
Outside the sphere, the electric field is given by 
\begin{align}
\tilde{\bm E}_\omega^{(1)}({\bm x})={\bm e}^0 \ee^{\ii{\bm k}\cdot{\bm x}}+ \sum_{LL'}\sum_{\zeta=\mathrm{M,N}}h_l(q_br)Y_L(\Omega){\bm P}_{LL}^\zeta \psi_{L'}^{\zeta >}, 
\end{align}
where $h_l$ is the spherical Hankel function of the first kind and integer order $l$. 
The coefficient $\psi_{L}^{\zeta >(<)}$ is determined via the S-matrix of the sphere \cite{bohren2008absorption}.

The electrostatic potential inside the sphere is given by 
\begin{align}
&\phi_\mathrm{DC}^{(2)}({\bm x})= \sum_{L}a_Lr^lY_L(\Omega) \nonumber \\
&\hskip30pt + \frac{1}{4\pi\epsilon_0\epsilon_\infty}\int _{r'\le R}d^3{\bm x}'\frac{1}{|{\bm x}-{\bm x}'|}\rho_\mathrm{DC}^{(2)}({\bm x}').
\end{align}
By imposing the open-circuit condition, we have 
\begin{align}
\partial_r\phi_\mathrm{DC}^{(2)}|_{r=R}&=\frac{\gamma}{\epsilon_0\omega_\mathrm{p}^2}({\bm i}_\mathrm{DC}^{(2)})_r|_{r=R}\nonumber \\
&=\sum_L N_LY_L(\Omega). 
\end{align}
The coefficient $a_L$ is determined as 
\begin{align}
&a_LlR^{l-1} = N_L + \frac{Q_L}{\epsilon_0\epsilon_\infty} \frac{l+1}{2l+1}R^{-l-2},\\
&Q_L=\int \dd^3{\bm x} r^l Y_L^*(\Omega)\rho_\mathrm{DC}^{(2)}({\bm x}).
\end{align}

The potential outside the sphere is simply given by 
\begin{align}
&\phi_\mathrm{DC}^{(2)}({\bm x})= \sum_{L}b_Lr^{-l-1}Y_L(\Omega), \\
&b_L=a_LR^{2l+1}+\frac{1}{\epsilon_0\epsilon_\infty}\frac{Q_L}{2l+1},
\end{align}
where coefficients $b_L$ are determined by the continuity of the potential itself.

The Maxwell boundary condition gives rise to the surface charge density $\sigma(\Omega)$ as 
\begin{align}
&\sigma(\Omega)=\sum_L\sigma_L Y_L(\Omega),\\
&\sigma_L=\epsilon_0(\epsilon_\infty l + \epsilon_b(l+1))a_LR^{l-1} \nonumber\\
&\hskip20pt -\frac{l+1}{2l+1}\left(1-\frac{\epsilon_b}{\epsilon_\infty}\right)Q_LR^{-l-2}. 
\end{align}
The $l=0$ component is rewritten as 
\begin{align}
&Q + Q_\mathrm{s}=\sqrt{4\pi}\epsilon_0\epsilon_bR\left(a_0+\frac{Q_0}{\epsilon_0\epsilon_\infty R}\right), 
\end{align}
where 
\begin{align}
Q = \sqrt{4\pi}Q_0,\quad Q_\mathrm{s}=\sqrt{4\pi}R^2\sigma_0, 
\end{align}
are the bulk and surface charges, respectively. 
We assume the charge neutrality $Q+Q_\mathrm{s}=0$ in the second-order nonlinear response, so that we have 
\begin{align}
0=a_0+\frac{Q_0}{\epsilon_0\epsilon_\infty R}=\frac{b_0}{R}.
\end{align}

Let us consider the photovoltage $V$ between the north and south poles of the sphere. The photovoltage is given by 
\begin{align}
V&=\phi_\mathrm{DC}^{(2)}(R\hat{z})-\phi_\mathrm{DC}^{(2)}(-R\hat{z}) \nonumber \\
&=\sum_{l=\mathrm{odd},m=0}\sqrt{\frac{2l+1}{4\pi}}\left(2a_LR^l+\frac{2}{\epsilon_0\epsilon_\infty}\frac{Q_L}{(2l+1)R^{l+1}}\right).
\end{align}  
  
\begin{figure}
	\centering
\includegraphics[width=0.45\textwidth]{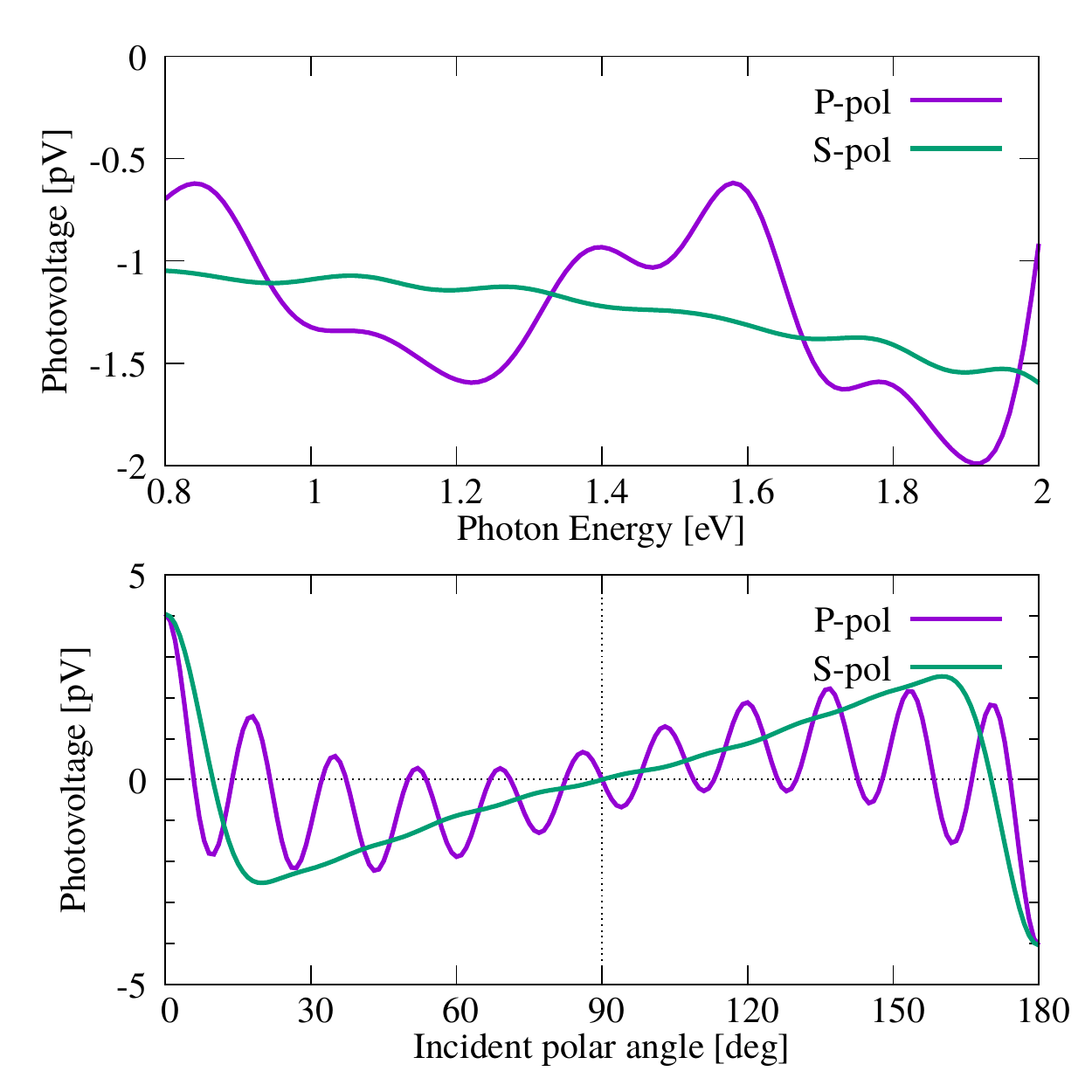}
	\caption{(upper panel) The photovoltage spectrum in an isolated Au sphere as a function of frequency of the incident plane-wave light.  The incident angle is fixed as $\theta_\mathrm{inc}=45^\circ$. (lower panel) The incident-angle dependence of the photovoltage at a dip  frequency of $\hbar\omega$=1.91 [eV] of the P-polarization. The sphere has radius $R=1$ [$\mu$m] and the medium outside the sphere is air ($\epsilon_b=1$).
    The flux density of the incident light is 500 [W/cm${}^2$]. The P and S polarizations are referred to the electric field polarization  parallel and perpendicular to the incident plane, respectively. The plane is formed by the incident wave vector and the $z$ axis. }
	\label{fig:test18or}
\end{figure}

Figure \ref{fig:test18or} shows the frequency spectrum and polar-angle dependence of the photovoltage of an isolated Au sphere. 
As in the cylinder case, the absolute values of the photovoltage are much lower than in the slab case. In the spherical system, the propagating surface plasmon is absent. Instead, particle-plasmon resonances can take place.  Within the simple Drude response Eq. (\ref{Eq_Drude}), the resonance frequencies are  given by  
\begin{align}
\omega=\omega_\mathrm{p}\sqrt{\frac{l}{l\epsilon_\infty+(l+1)\epsilon_b}}, 
\end{align}
for angular momentum $l$, under the non-retarded approximation. 
In our case of Au, however, the resonance frequencies are above the interband-transition threshold, where the simple Drude formula fails to reproduce the experimentally-measured dielectric constant of Au  \cite{west2010searching}. 
Therefore, the particle plasmon resonances are not efficiently described within the present formalism.

As for the angular dependence, we can see clearly the anti-symmetric dependence with respect to $\theta_\mathrm{inc}=90^\circ$, at which the photovoltage vanishes. The photovoltage exhibits the maxima and minima when the incident light is coming along the $z$ axis ($\theta_\mathrm{inc}=0^\circ,180^\circ$). If the incident wave vector is tilted from the $z$ axis, the photovoltage for the P-polarized light oscillates with the incident angle, in a striking contrast to that for the S-polarization.

\begin{figure}
	\centering
\includegraphics[width=0.45\textwidth]{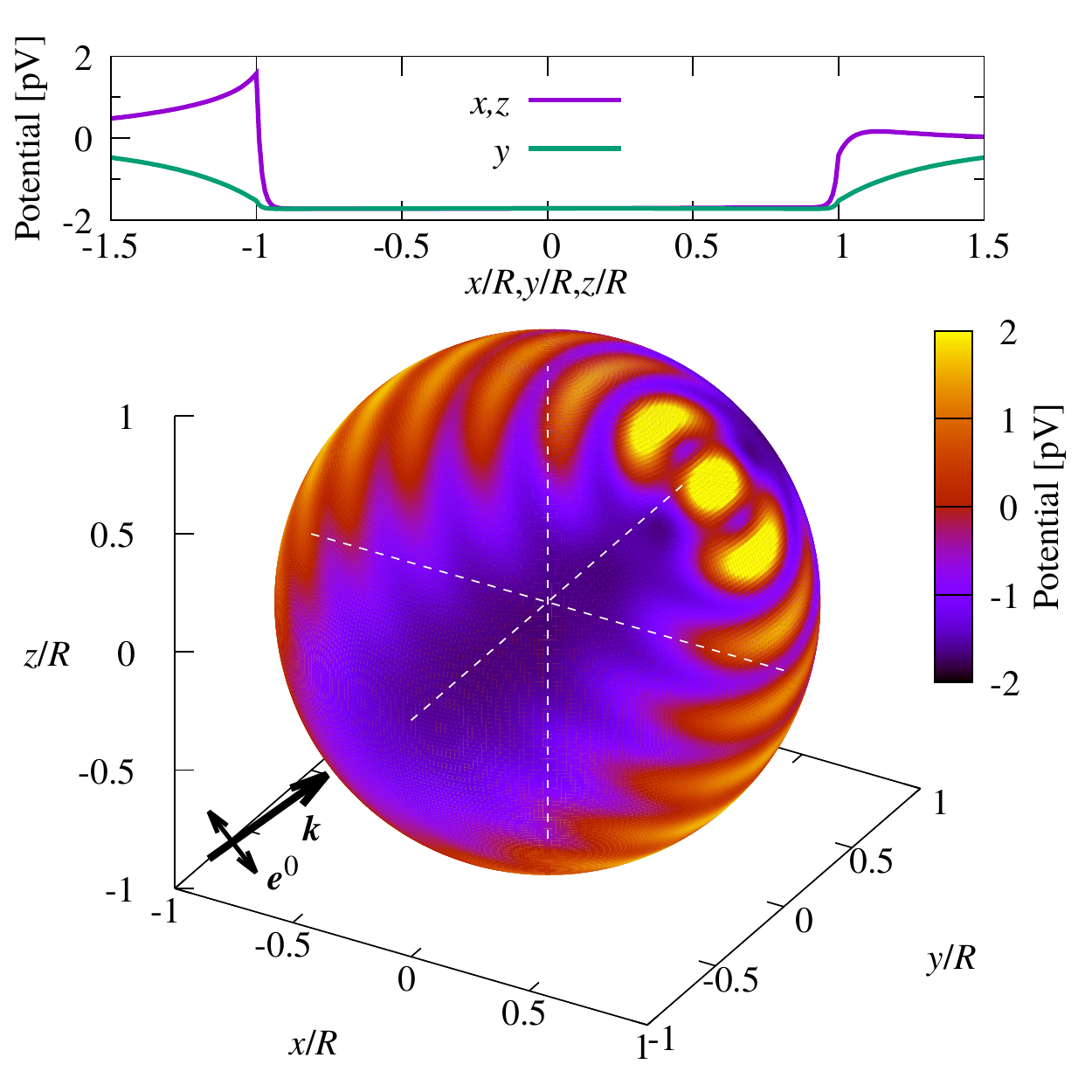}
	\caption{(upper panel) The electrostatic potential along the $x$, $y$ , and $z$ axes depicted with dashed lines in the lower panel.  (lower panel) The potential distribution on the sphere surface.
		The P-polarized plane-wave light of $\hbar\omega$=1.91 [eV] is coming with incident angle of $(\theta_\mathrm{inc},\varphi_\mathrm{inc})=(45^\circ,0^\circ)$.
The other parameters are the same as in Fig \ref{fig:test18or}. 	
}
	\label{fig:test18pw191pot}
\end{figure}

A typical electrostatic-potential distribution is shown in Fig.~\ref{fig:test18pw191pot} for the P-polarized incident light. 
On the sphere surface, the potential exhibits an marked fringe. 
This fringe has the same origin as in the angular dependence of Fig. \ref{fig:test18or}. That is, changing the incident angle is equivalent to changing the probe positions from the north and south poles simultaneously, picking up the spatial fringe of the potential. 
We found that at this incident light, the multipole components $b_L$ of angular momenta around $l=2$ and 21 have dominating contributions. Since the $l=2$ spherical harmonics has rather smooth dependence on the solid angle, the spherical harmonics around $l=21$ is crucial in the spatial fringe of the potential.         
Inside the sphere, the potential is almost constant. 
Outside the sphere, the potential becomes zero at $r\to\infty$ faster than $1/r$, because the monopole coefficient $b_0$ vanishes by the cancellation between the bulk and surface charges.

\section{Summary and discussion}
In summary, we have formulated the electrostatic theory of the optical rectification in arbitrary geometries of metals, using the hydrodynamical approach to metal carries within the local response approximation.  We then have applied the theory to metallic planar interfaces, slabs, cylinders, and spheres.  
The photovoltage and photocurrent spectra, together with the electrostatic potential distributions are demonstrated for these systems made of gold. 

We found that the photovoltage and photocurrent are strongly enhanced by a plasmonic resonance under the Kretschmann geometry of the Au slabs. However, the plasmonic resonance are hidden particularly in the Au sphere, because the resonance frequencies in the simple Drude response, which is assumed in the hydrodynamics approach, are above the interband-transition threshold. There, the simple Drude fitting is no longer available.

In the present formulation, we have neglected possible nonlocal effects 
and have focused on possible geometrical effects taking account of metal boundaries. 
Although the analytic approach presented in the paper is available only in the limited geometries, the numerical approach with integrated solvers of the Maxwell and Poisson equations can handle the systems with arbitrary finite geometries. In addition, if we approximate a complex metal geometry by a composite of metallic slabs, cylinders, and spheres, 
the present analytic formulation can be applied by taking account of the multiple scattering among such geometrical ingredients. 

There are several possibilities of extending the present electrostatic theory of the OR. Including nonlocal responses and/or quantum spill-out effects and/or symmetry breaking effects by some optical anisotropy are important directions of the extension. 
We believe that many questions remained in the OR of metallic nanostructures are finally resolved after pursuing these directions.
We hope this paper stimulates further investigation of the metallic OR along these lines.





\begin{acknowledgments}
The author would like to thank Prof. Jimmy Xu for valuable discussion and suggestions.  He also likes to thank Dr. P. Moroshkin and Dr. R. Osgood III for various inputs in a related collaboration, which motivated the author to pursue the present theoretical study.   
\end{acknowledgments}


%

\end{document}